\documentclass[a4paper]{article}
\usepackage{amsmath,amssymb,amscd,amsfonts,latexsym,mathrsfs}
\parskip=\medskipamount

\newcommand{\sect}[1]{\setcounter{equation}{0}\section{#1}}

\newcommand{\vs}[1]{\rule[- #1 mm]{0mm}{#1 mm}}
\newcommand{\hs}[1]{\hspace{#1 mm}}
\newcommand{\lbl}[1]{\label{eq:#1}}
\newcommand{\rf}[1]{(\ref{eq:#1})}
\newcommand{\nn}{\nonumber}

\newcommand{\ee}{\vs{2}\end{equation}}

\newcommand{\bea}{\begin{eqnarray}}
\newcommand{\ena}{\end{eqnarray}}
\newcommand{\nnbea}{\begin{eqnarray*}}
\newcommand{\nnena}{\end{eqnarray*}}

\newcommand{\tld}{\widetilde}
\newcommand{\lra}{\ \longrightarrow\ }

\newcommand{\ovl}[1]{\overline{#1}}

\newcommand{\sm}[2]{\textstyle{\frac{#1}{#2}}\displaystyle}

\newcommand{\dps}{\displaystyle}

\newcommand{\bz}{{\ovl{z}}}

\newcommand{\bc}{{\ovl{c}}}

\newcommand{\lam}{\lambda}

\newcommand{\zbz}{(z,\ovl{z})}

\newcommand{\zbzr}{(z,\ovl{z}|r)}

\newcommand{\zbzs}{(z,\ovl{z}|s)}

\newcommand{\cR}{{\cal R }}
\newcommand{\cW}{{\cal W }}

\newcommand{\cWs}{{\cal W }_{s}}

\newcommand{\cU}{{\cal U }}

\newcommand{\cK}{{\cal K }}

\newcommand{\cF}{{\cal F }}
\newcommand{\cC}{{\cal C }}

\newcommand{\cT}{{\cal T }}

\newcommand{\cM}{{\cal M }}
\newcommand{\cS}{{\cal S }}

\newcommand{\cB}{{\cal B }}
\newcommand{\cX}{{\cal X }}

\newcommand{\cQ}{{{\cal Q }}}

\newcommand{\cJ}{{\cal J }}

\newcommand{\prt}{\partial}
\newcommand{\prtz}{\partial_z}

\newcommand{\prtbz}{\ovl{\partial}_{\ovl{z}}}

\newcommand{\bprt}{\ovl{\partial}}

\newcommand{\wbw}{(w,\ovl{w})}

\newcommand{\wbws}{(w,\ovl{w}|s)}

\newcommand{\z}{(z)}

\newtheorem{Conclusion}{Conclusion}[section]

\newtheorem{Statement}{Theorem}[section]

\newtheorem{Proof}{Proof}
\newtheorem{Properties}{Properties}
\newtheorem{Remark}{Remark}[section]

\newtheorem{Conjecture}{Conjecture}[section]

\begin{document}

\title{\bf Large Chiral Diffeomorphisms on Riemann Surfaces and $\cW$-algebras}

\vskip -1cm

\author{ {\sc G. BANDELLONI} $^a$
 and {\sc S. LAZZARINI} $^b$\\[6mm]
$^a$ Dipartimento di Fisica
dell'Universit\`a di Genova,\\
 Via Dodecaneso 33, I-16146 GENOVA-Italy\\
 and \\
Istituto Nazionale di Fisica Nucleare, INFN, Sezione di Genova\\
via Dodecaneso 33, I-16146 GENOVA Italy\\
e-mail : {\tt beppe@genova.infn.it}\\[4mm]
$^b$ Universit\'e de la M\'editerran\'ee, Aix-Marseille II\\
Centre de Physique Th\'eorique
\footnote{Unit\'e Mixte de Recherche
(UMR 6207) du CNRS et des Universit\'es Aix-Marseille I, Aix-Marseille II et
de l'Universit\'e du Sud Toulon-Var, affili\'ee \`a la
FRUMAM (F\'ed\'eration de Recherche 2291).}\ , Case postale 907,\\ 
 F--13288 Marseille Cedex 9, France\\
e-mail : {\tt sel@cpt.univ-mrs.fr} }
\maketitle


\centerline{\bf Abstract} 

The diffeomorphism action lifted on truncated (chiral) Taylor expansion of a
complex scalar field over a Riemann surface is presented in the paper under
the name of large diffeomorphisms. After an heuristic approach, we show
how a linear truncation in the Taylor expansion can generate an algebra of
symmetry characterized by some structure functions.  
Such a linear truncation is explicitly realized by introducing the
notion of Forsyth frame over the Riemann surface  with the
help of a conformally covariant algebraic differential equation.
The large chiral diffeomorphism action is then implemented  through a
B.R.S. formulation (for a given order of truncation)  leading  to a
more algebraic set up.  In this context the ghost fields behave as
holomorphically covariant jets.  Subsequently, the link with the so
called  $\cW$-algebras  is made explicit once the ghost parameters are
turned from jets into tensorial ghost ones. We give a  general
solution with the help of the structure functions pertaining to all
the possible truncations lower or equal to the given order. This
provides another contribution to the relationship between KdV flows
and $\cW$-diffeomorphims.

\indent

\noindent
PACS number :  11.25 hf

\noindent
Keywords : Conformal field theory- $\cW$-algebras.

\vfill

\noindent
CPT--2005/P.016

\newpage

\section{Introduction}

The notion of symmetry gives a structure to spacetime  (or
configuration space)
and/or internal spaces of the model under consideration, in the sense
that the former is closely related to a geometrical setup.

At the infinitesimal level, the concept of algebra turns out to be very
useful.  It generally gives rise to a solution when the linearity is
fulfilled. In the non linear situation, however, we have sometimes
to explore further behind the first infinitesimal  transformation step.
 
The development of non linear sciences has been supported these needs
\cite{deBoer:1996nu}, and many basic   physical systems were described
by non linear  extensions of algebras. This is the case for integrable
systems, two dimensional conformal   models (with application to Strings,
Gravity or Solid State Physics, see e.g. \cite{Tjin:1993dk}).

Particular interest has been devoted to the so-called $\cW$-algebras
\cite{Zamolodchikov:1985wn}  which come out  from different
principles \cite{Drinfeld:1984qv,Bilal:1991wn,DiFrancesco:1991qr} and
the question of their geometric origin remains still unclear or
unsatisfactory despite the various attempts given by
\cite{Bilal:1991wn,Zuc93,Gov95,AF95,Gov96,Bandelloni:1999et}. 
   
In particular, the relationship between a conformally covariant $s$-th order
differential equation 
\begin{eqnarray*}
  \bigg( \prt_{(s)} + a^{(s)}_{(2)}\zbz \prt_{(s-2)} + \cdots +
  a^{(s)}_{(s)}\zbz \bigg) f\zbz = 0\ ,
\end{eqnarray*}
over a generic Riemann surface and some of the so-called
$\cW$-algebra has been well established since fifteen years by
\cite{DiFrancesco:1991qr}. The differential equations can be thought
of as equations of motion for some matter fields.  The former can be
also derived as
vanishing covariant derivative condition \cite{Garajeu:1995jn}, and
in some extent, more general $\cW$-algebra as the Bershadsky one
\cite{Ber91}, can be related to a system of conformally covariant
coupled differential equations. 

Focussing here only on one $s$-th order differential equation, one may
either work directly with the solutions or with uniformizing
coordinates considered as ratios of  linear independent solutions
\cite{Forsyth59,Gov95}. It turns out to be a matter of taste of 
working either with homogeneous coordinates or inhomogeneous one on
$\mathbb{C}P^{s-1}$. However, in his
textbook \cite{Forsyth59}, Forsyth uses rather  inhomogeneous
coordinates in order to get some differential invariants.  The
inhomogeneous coordinates have to satisfy a $s+1$-th order equations.
Accordingly, our point of view proposed in \cite{BaLa04} traced back in the
literature, and found out an unexpected origin in old past 
\cite{Forsyth59} related to some algebraic type differential
equations, once covariantly formulated over a generic Riemann surface.
Indeed the definition of general well-defined differential operators
 over a (two dimensional)  Riemann Surface requires some care
\cite{Bol}, and puts forward a deep insight on the links  between
  covariance required by physical considerations and projective
  geometry. Related studies of 
projectively invariant differential operators as well differential
invariants can be found in \cite{Wil}.

In \cite{BaLa04} the notion of Laguerre-Forsyth frames were
promoted. In order to avoid a possible confusion
with the named Laguerre-Forsyth form of a conformally covariant differential
equation \cite{Gov95}, in the present paper, we adopt the name Forsyth frames.
We have simply in mind the ideas of, firstly,
pursuing further ahead the method given by Forsyth in
\cite{Forsyth59}, and secondly, dealing with scalar coordinates
considered as solutions of generalized Beltrami equations (see
e.g. \cite{Gov95} or \cite{Bandelloni:2002zg} appendix C2)
just about the approach given in \cite{Gov95,Gov96}.
Our motivation for using the inhomogeneous coordinates rests on the
fact that they seem to be more natural for constructing projective invariants.
Projective action of $SL(s,\mathbb{C})$ on $\mathbb{C} P^{s-1}$. In
general \cite{GlHB97},  the smooth coefficients in the above $s$-th
order covariant differential equation  have been proven to be
projectively invariant. 
Inhomogeneous coordinates are the local coordinates for projective
curves in $\mathbb{C} P^{s-1}$ on which there is a symplectic structure related
to the Gelfand-Dickey brackets \cite{GO94,GlHB97}.
Moreover, inhomogeneous coordinates offer the
possibility to work with scalar fields instead of densities.

In \cite{BaLa04}, these Forsyth frames were thus made of coordinate scalar
fields on some finite 
dimensional target space and constructed from solutions of a finite
order holomorphically covariant linear differential equation over a Riemann
surface of the above type.  At the quantum level, these scalar fields
surprisingly 
gained already at the one-loop level a non commutative character. This
phenomenon gives a quantum origin to a non commutative structure on the
target space. The non-commutativity came out by anomaly cancellation as
a nonlocal effect.

\medskip

The basic novelties of these Forsyth frames lie in their non trivial
properties under  differentiations, which allow to expand beyond the
first order the holomorphic  reparametrization process still
maintaining the algebra closure property. Thus investigate this type of
extended algebra appears as a necessity, since this ``new''
structure encapsulates the (general) covariance laws.

\medskip

In this paper, the construction of Forsyth frames  is
proposed in great details and it strongly relies on a symmetry
principle. The latter will be systematically studied in the algebraic
BRS language more suitable for a possible field theoretical treatment of the
model.

The paper is organized as follows.
Section \ref{Motiv} goes over some motivations for this extended
notion of symmetry, beyond linearity.  It requires the use of higher
order derivatives and the closure of the algebra obliges to introduce
new fields which will play the role of structure functions for the
(non linear) reduced symmetry algebra.  Section \ref{FL} is devoted to the
definition of the Forsyth frames and to a deep study of their
properties.  In Section \ref{BRS}, a convenient BRS approach is
presented for the algebra of symmetry. It will be useful for improving in
particular the covariance laws which come out from these properties.
Section \ref{Covar} treats the covariance under holomorphic change
charts of the algebra   elements. Non tensorial structures (jet) come
out, in particular under the form of jet-BRS ghosts which are harmful
for physical considerations since covariant quantities are required. A
rather tricky link with tensorial ghosts is presented  as a change of
basis of generators.  Section \ref{W}  gives a way (using all the
differential properties of the Forsyth frames, including the
ones  of the sub-frames encapsulated into the maximal one) to
decompose into tensors the jet-BRS ghosts.  The process is obviously
defined up to a tensorial rescaling.  Some detailed examples are given
and illustrate the striking property of cancelling out the effects of
all the sub-frames, in favor to a standard presentation of the
$\cW$-algebra structure. Exploiting some known results in the literature
\cite{Gov96} allows to clarify the algorithm.
 We conclude in section \ref{concl}.

\sect{Motivations} \label{Motiv}

The issue of finding the most general expression of a spacetime
symmetry can find an answer in the concept of generalized frames,
perhaps rather of that of prolongated frames whose local expression is
obtained by successive derivations \cite{Koba95}. In order to consider
such objects,  let us first think of the Fock space of some smooth
complex scalar field $Z$ defined over a given  Riemann surface
$\Sigma$, endowed with a complex analytic (holomorphic) structure;
this requires the use of local complex coordinates $\zbz$. Smoothness
of the complex scalar field $Z=Z\zbz$ is understood with respect to
the differential  structure on $\Sigma$ with which the complex
structure is subordinated to.  From the locality principle, recall
that the Fock space for $Z$ is generated by all the $z$ and $\bz$
derivatives of any order considered as independent monomials.

Consider now the infinitesimal action of
smooth diffeomorphisms on $\Sigma$ which is usually expressed by means of the
Lie derivative $L_\xi Z = (\xi\zbz \prt + \ovl{\xi}\zbz \bprt) Z$. 

With respect to the complex structure we shall be concerned with the
so-called ``chiral'' diffeomorphisms acting on the complex scalar
field $Z$ which consist in separating the Lie derivative according to
$z$-derivative, so that $Z \lra Z + \xi \prt Z$; there is the complex
conjugate expression as well. Denoting in the Fock space the various
$z$-derivatives of order $\ell$ by $\prt_{(\ell)} Z$, $\ell
=1,2,\dots$ one wants to consider a fully chiral local variation for
the complex  scalar field $Z$ (going thus beyond the first order) of
the following type
\begin{eqnarray}
  (\delta Z)\zbz = \sum_{\ell\geq 1}
\gamma^{(\ell)}\zbz\, \prt_{(\ell)} Z\zbz,
\lbl{lardiff}
\end{eqnarray}
together with the complex conjugate expression, and into which all the
$z$-derivatives appear.

Hence, constructing a local field theory over $\mathbb{C}$ amounts to
working a priori on local functionals in the various $z$-derivatives
of $Z$. So to speak, the physical model is built over the infinite
$z$-frames which is locally represented by $(Z\zbz,\prt
Z\zbz,\prt_{(2)}Z\zbz,\dots)$, which reproduces the ``chiral'' Taylor expansion
of the field $Z$ (i.e. only with respect to the $z$-coordinate) at the
point $\zbz$ of $\Sigma$, in other words, the infinite jet of $Z$.

\smallskip

What is called large chiral diffeomorphisms in the paper, is the lifted action
of usual local chiral diffeomorphisms on $\mathbb{C}$ to the infinite jet
space $J^\infty(\mathbb{C}, \mathbb{C})$, i.e. on the $z$-Taylor expansion.
The former are viewed as transformations acting on the complex scalar field
$Z$ itself and they require, on the one hand, infinitely many local
parameters $\gamma^{(\ell)}$ of conformal type $(-\ell,0)$ which
generalize vector fields of type $(-1,0)$, and, on 
the other hand any higher order derivatives of $Z$.

\medskip

Accordingly, it is called a large chiral diffeomorphism symmetry the
invariance of some observables or functional on $J^\infty(\mathbb{C},
\mathbb{C})$ under the transformation \rf{lardiff}.  Translating this
problem of symmetry on the space of local functionals in the infinite
jet of the scalar field $Z$,
 \begin{eqnarray}
  \delta Z\zbz = \left(\sum_{\ell\geq 1} \int_\mathbb{C} d\ovl{w}\wedge dw\
  \gamma^{(\ell)}\wbw \cW_{(\ell)}\wbw \right) Z\zbz,
\lbl{deltaZ}
\end{eqnarray}
amounts to introducing local Ward operators associated to the local
parameters~$\gamma^{(\ell)}$ 
 \begin{eqnarray}
\cW_{(\ell)}\zbz = \prt_{(\ell)} Z\zbz \frac{\delta} {\delta Z\zbz},\qquad
\mbox{for } \ell\geq 1,
\lbl{Ward0}
\end{eqnarray}
which generate an infinite dimensional Lie algebra.
But note that the only Lie sub-algebra is for $\ell=1$ of which the
bracket (as tensorial product of distributions) writes
\begin{eqnarray}
\Big[\cW_{(1)}\zbz,\cW_{(1)}\wbw\Big] &=& \cW_{(1)}\zbz
  \prt_w\delta(w-z) -  \cW_{(1)}\wbw\prt_z \delta(z-w)\nn\\
  &=& \cW_{(1)}\wbw\prt_w\delta(w-z) - \cW_{(1)}\zbz\prt_z \delta(z-w)
\end{eqnarray}
and translates by duality the Lie algebra structure of vector fields to the
functional level, namely, if $W_1(\xi) = \Big\langle \cW_1 \, |\, \xi
  \Big\rangle$ then
\begin{eqnarray}
\Big[W_1(\xi), W_1(\eta)\Big] &=& \Big\langle
\Big[\cW_{(1)}\zbz,\cW_{(1)}\wbw\Big]\, \Big|\, \xi\zbz \eta\wbw
\Big\rangle \nn\\
&=&
 \Big\langle \cW_1 \, \Big|\, \eta\prt\xi - \xi\prt\eta \Big\rangle =
\Big\langle \cW_1 \, \Big|\, [\eta,\xi] \Big\rangle = W_1([\eta,\xi])
\end{eqnarray}
and thus reproduces the bracket between the parameters of usual conformal
 transformations.
 (The pairing $\langle\, |\,\rangle$ stands for the functional
evaluation of distributions). In full generality, the brackets for
$k,\ell\geq1$ read
\begin{eqnarray}
  \Big[\cW_{(k)}\zbz,\cW_{(\ell)}\wbw\Big] &=& \sum_{m=1}^\ell
  \left(\!\!
\begin{array}{c}
\ell \\ m
  \end{array} \!\!\right)
\prt_w^m\delta(w-z)\cW_{(k+\ell-m)}\wbw \nn\\
&&\qquad
- \sum_{m=1}^k
  \left(\!\!
\begin{array}{c}
k \\ m
  \end{array} \!\!\right)
\prt_z^m\delta(z-w)\cW_{(k+\ell-m)}\zbz
\lbl{bracket0}
\end{eqnarray}
and close onto subspaces generated by
$\{\cW_{(u)}\}_{u=\mbox{\small min}(k,\ell)}^{k+\ell-1}$ and leads to
introducing higher order generators at each step.  Note that the
bracket \rf{bracket0} fulfills the Jacobi identity. Moreover for
arbitrary $k$ and $\ell=1$ one obtains by duality
\begin{eqnarray}
\Big\langle
\Big[\cW_{(k)}\zbz,\cW_{(1)}\wbw\Big]\, \Big|\, \xi^{(k)}\zbz \eta^{(1)}\wbw
\Big\rangle &=&
\Big\langle \cW_{(k)}\, \Big|\, \eta^{(1)}\prt\xi^{(k)} - k\, \prt\eta^{(1)}
\xi^{(k)} \Big\rangle\nn\\
&& -\ \sum_{m=2}^{k\geq 2} \begin{pmatrix} k \\ m \\
\end{pmatrix} \Big\langle \cW_{(k-m+1)}\, \Big|\, \xi^{(k)} \prt^m
\eta^{(1)} \Big\rangle
\lbl{Wconfor}
 \end{eqnarray}
where the first smearing bracket on the right hand side shows that the
conformal transformations $\cW_{(1)}$ preserve the $\cW_{(k)}$
transformations up to lower orders and defines a covariant bracket
$[\eta^{(1)},\xi^{(k)}]^{(k)}=\eta^{(1)}\prt\xi^{(k)} - k\,
\prt\eta^{(1)} \xi^{(k)}$
showing that the parameter $\xi^{(k)}$ carries a conformal weight $(-k,0)$.

\medskip

The closure onto a finite dimensional Lie sub-algebra
for $\ell>1$, can be obtained by a truncation in the $z$-derivatives
of $Z\zbz$ at the some finite order, say $s-1$, ($s\geq
2$). Setting for $z$-derivatives of order greater than $s-1$, namely
for $m\geq s$, the following linear combinations
  \begin{eqnarray}
  \prt_{(m)} Z\zbz &=& \sum_{\ell=1}^{s-1 }\cR_{(m)}^{(\ell)}\zbz\,
  \prt_{(\ell)}Z\zbz
  \Longrightarrow
\cW_{(m)} \zbz = \sum_{\ell=1}^{s-1} \cR_{(m)}^{(\ell)}\zbz\, \cW_{(\ell)}\zbz
\lbl{Z_trunca}
  \end{eqnarray}
where the finite summation run over $\ell = 1,\dots,s-1$, thus the
immediate consequence is that the bracket \rf{bracket0} reduces to
\begin{eqnarray}
  \hskip -2cm \Big[\cW_{(k)}\zbz,\cW_{(\ell)}\wbw\Big] &=&
\sum_{u=1}^{s-1} \left\{ \sum_{m=1}^\ell
  \left(\!\!
\begin{array}{c}
\ell \\ m
\end{array} \!\!\right)
\prt_w^m \delta(w-z)\cR_{(k+\ell-m)}^{(u)}\wbw
\cW_{(u)}\wbw \right. \nn\\
&& \left. \qquad\quad
- \sum_{m=1}^k
  \left(\!\!
\begin{array}{c}
k \\ m
\end{array} \!\!\right)
\prt_z^m\delta(z-w)\cR_{(k+\ell-m)}^{(u)}\zbz
\cW_{(u)}\zbz \right\},
\lbl{bracket}
\end{eqnarray}
and closes onto the generators $\{\cW_{(u)}\}_{u=1}^{s-1}$. However,
the truncation spoils the Jacobi identity due to the restriction of
the order to the range $\ell = 1,\dots, s-1$. The Jacobi identity
(which guarantees the associativity of the Lie algebra) will be
restored by modifying the generators $\{\cW_{(u)}\}_{u=1}^{s-1}$  in
such a way to take into account the reduction coefficients
$\cR_{(v)}^{(u)}\zbz$ introduced in \rf{Z_trunca}. These coefficients
will play the role of  ``structure 
functions'' for the finite dimensional Lie algebra generated by the
modified generators $\{\tld\cW_{(u)}\}_{u=1}^{s-1}$. Owing to \rf{Z_trunca}, this is achieved
by computing, for $k\geq s$ and for $\ell = 1,\dots, s-1$, the difference
\begin{eqnarray}
\Big[\cW_{(k)}\zbz,\cW_{(\ell)}\wbw\Big] - \sum_{p=1}^{s-1}
    \cR^{(p)}_{(k)}\zbz  \Big[\cW_{(p)}\zbz,\cW_{(\ell)}\wbw\Big] &=:& \nn\\
&&\hskip -5cm \sum_{u=1}^{s-1} \Big[\cR^{(u)}_{(k)}\zbz
    ,\cF_{(\ell)}\left(\cR\wbw,\frac{\delta}{\delta\cR\wbw}\right)
    \Big]\cW_{(u)}\zbz, 
\lbl{diffward}
\end{eqnarray}
where $\cF_{(\ell)}\left(\cR\wbw,\frac{\delta}{\delta\cR\wbw}\right)$
is a functional differential polynomial in the $\cR$'s and insures the
modification of the generator, for each $\ell = 1,\dots,s-1$, according to
\begin{eqnarray}
\cW_{(\ell)}\wbw \lra \tld\cW_{(\ell)}\wbw = \prt_{(\ell)} Z\wbw
\frac{\delta} {\delta Z\wbw} +
\cF_{(\ell)}\left(\cR\wbw,\frac{\delta}{\delta\cR\wbw}\right) 
\end{eqnarray}
in view to fulfill the Jacobi identity. Furthermore, of course one has
for $\ell = 1,\dots,s-1$ 
\begin{eqnarray}
  \cR_{(k)}^{(\ell)}\zbz = \begin{cases} \delta_k^\ell & \mbox{if } 1\leq
  k \leq s-1, \\[2mm]
\cR_{(k)}^{(\ell)}\zbz & \mbox{if } k\geq s,
\end{cases}
\lbl{R[s]}
\end{eqnarray}
and therefore the functional operator $\cF_{(\ell)}$ must contain
functional derivatives with respect to the structure functions
$\cR_{(k)}^{(p)}$ for $k\geq s$ only. Thus inserting twice the
brackets \rf{bracket} into the right hand side of \rf{diffward}, a
direct comparison with the left hand side of \rf{diffward} amounts, on
the one hand, to the vanishing of the coefficient terms of the
$\cW_{(u)}\wbw$'s. This gives rise to some compatibility conditions
that must be fulfilled by the structure functions, namely,
\begin{eqnarray}
\cR^{(u)}_{(k+n)} \wbw = \sum_{p=1}^{s-1} \sum_{j=0}^n \binom{n}{j}
\prt_{(j)} \cR_{(k)}^{(p)}\wbw \cR_{(p+n-j)}^{(u)}\wbw,\qquad \mbox{
  for } n=0,\dots,\ell \mbox{ and } k\geq s. 
\lbl{prtRJacobi}
\end{eqnarray}
If $k$ is taken to be lower or equal to $s-1$ then \rf{prtRJacobi}
restricts to $\cR^{(u)}_{(k+n)} = \cR^{(u)}_{(k+n)}$, since
$\cR_{(k)}^{(p)}=\delta_k^p$. 

On the other hand, the coefficient term of $\cW_{(u)}\zbz$ provides
the functional differential operator
\begin{eqnarray}
        \cF_{(\ell)}\left(\cR\wbw,\frac{\delta}{\delta\cR\wbw}\right)
        &=& \sum_{i\geq s} \sum_{j=1}^{s-1} \left\{  \sum_{m=0}^i
        \binom{i}{m} (-1)^m \prt_{(m)}\left(
        \cR_{(i+\ell-m)}^{(j)}\wbw
        \frac{\delta}{\delta\cR_{(i)}^{(j)}\wbw}\right) \right. \nn\\
        && \left. \hskip -2cm -\ \sum_{p=1}^{s-1}\sum_{q=0}^p
        \binom{p}{q}(-1)^q \prt_{(q)}\left( \cR_{(i)}^{(p)}\wbw
        \cR_{(p+\ell-q)}^{(j)}\wbw
        \frac{\delta}{\delta\cR_{(i)}^{(j)}\wbw} \right) \right\}.
        \lbl{Ward1}
\end{eqnarray}
Therefore, in addition to the scalar field $Z$, the structure
functions $\cR$'s come as new fields to be taken into account in the
theory.  Their variation is obtained to be
\begin{eqnarray}
  \delta \cR_{(n)}^{(p)}\zbz &=& \left(
  \int_\mathbb{C} d\ovl{w}\wedge dw\ \left[ \sum_{\ell = 1}^{s-1} 
\gamma^{(\ell)}\wbw + \sum_{u\geq s} \gamma^{(u)}\wbw
  \cR^{(\ell)}_{(u)}\wbw \right] \tld\cW_{(\ell)}\wbw \right)
  \cR_{(n)}^{(p)}\zbz\nn\\[2mm]
&=& \sum_{\ell = 1}^{s-1} \left[ \sum_{m=0}^n \binom{n}{m} \prt_{(m)}
  \left( \gamma^{(\ell)} + \sum_{u\geq s}\gamma^{(u)}
  \cR^{(\ell)}_{(u)} \right) \cR_{(n + \ell - m)}^{(p)} \right. \nn\\[-2mm]
&&\lbl{deltaR} \\[-2mm]
&&\qquad \left. -\ \sum_{j = 1}^{s-1} \cR^{(j)}_{(n)} 
\sum_{q=0}^j \binom{j}{q} \prt_{(q)} \left( \gamma^{(\ell)} +
  \sum_{u\geq s}\gamma^{(u)} 
  \cR^{(\ell)}_{(u)} \right) \cR_{(j + \ell - q)}^{(p)} \right]\zbz, \nn
\end{eqnarray}
while the variation \rf{deltaZ} for $Z$  rewrites
\begin{eqnarray}
  \delta Z\zbz &=& \left( \int_\mathbb{C} d\ovl{w}\wedge dw\ \left[
      \sum_{\ell = 1}^{s-1}  
\gamma^{(\ell)}\wbw + \sum_{u\geq s} \gamma^{(u)}\wbw
  \cR^{(\ell)}_{(u)}\wbw \right] \tld\cW_{(\ell)}\wbw \right)  Z\zbz.
\lbl{deltaZbis}
\end{eqnarray}
To this change $\cW_{(\ell)}\lra\tld\cW_{(\ell)}$ (for $\ell=1,\dots,s-1$)
of generators there corresponds a reduction from an infinite number to
a finite number of local parameters
\begin{eqnarray}
\gamma^{(m)} \lra \Gamma^{(\ell)} = \sum_{m\geq 1} \gamma^{(m)}
\cR_{(m)}^{(\ell)}, \qquad
\ell=1,\dots,s-1
\lbl{Gamma}
\end{eqnarray}
as suggested by both the variations \rf{deltaR} and
\rf{deltaZbis}. The $s-1$ local 
parameters $ \Gamma^{(\ell)}$ secure the fact that the $s-1$ generators
$\tld\cW_{(\ell)}$ fulfill  indeed the algebra \rf{bracket}. In short,
this leads to a reduction of the symmetry algebra, and \rf{lardiff}
reduces to the variation
\begin{eqnarray}
  (\delta Z)\zbz = \sum_{\ell =1}^{s-1}
\Gamma^{(\ell)}\zbz\, \prt_{(\ell)} Z\zbz.
\lbl{lardiff_trunc}
\end{eqnarray}

By duality the following brackets $[\ ,\,]^{(u)}$ corresponding to the
generators $\tld\cW_{(u)}$ are found to be
\begin{eqnarray}
  [\eta^{(\ell)},\xi^{(k)}]^{(u)} = \sum_{m=0}^{\ell-1}
\begin{pmatrix} \ell \\ m \\ \end{pmatrix}
\cR^{(u)}_{(k+m)}\, \eta^{(\ell)} \prt_{(\ell-m)} \xi^{(k)}
- \sum_{m=0}^{k-1}
\begin{pmatrix} k \\ m \\ \end{pmatrix}
\cR^{(u)}_{(\ell+m)}\, \xi^{(k)} \prt_{(k-m)} \eta^{(\ell)}.
\end{eqnarray}
These brackets are involved in the defining Poisson brackets
for $\cW$-algebras \cite{Zamolodchikov:1985wn,Watts95,Watts98}. 
This leads to the  
\begin{Conclusion}
The  realization of large diffeomorphism algebra Eq.\rf{bracket0} requires
 the definition of frames which verify the truncation property
 Eq.\rf{Z_trunca} which realizes a  
 derivative order reduction (D.O.R). So the structure functions
 $\cR^{(u)}_{(v)}\zbz$ uniquely define the properties of the algebra. 
\end{Conclusion}

The problem we are after is twofold.  First, due to the presence of
higher order derivatives which carry a non tensorial nature (jets), one
wants to  perform the construction in a well defined  way, in the
sense that this local symmetry has indeed a global meaning over the
Riemann surface $\Sigma$. That is, constructing a field theory over the
coframes $J^\infty(\Sigma,\mathbb{C})$. Second, find the appropriate
generators for the symmetry algebra, (jets or tensors), which give
rise to covariant quantities over the Riemann surface these quantities being
constructed from covariant differential operators, covariant in the
sense to be holomorphically well defined on $\Sigma$. This would
correspond to a change of generators $\{\tld\cW_{(u)}\}_{u=1}^{s-1}$ in
order to get a presentation of the Poisson $\cW$-algebras which are no
longer Lie algebras. This means in particular that the lower orders in the
brackets \rf{Wconfor} should not be present any more \cite{Watts98}.

This second goal requires in fact to work with a finite number of
$s-1$ complex scalar
fields $Z$ instead of one only. Therefore, one is led to consider the
jet space $J^\infty(\Sigma,\mathbb{C}^{s-1})$ on which local
diffeomorphisms of $\mathbb{C}^{s-1}$ stabilizing the target point
$(Z^{(1)},\dots,Z^{(s-1)})$ are lifted by jet composition and act linearly. Presently, a truncation in the order of the jet can be implemented by means of relations given by some PDE's.
The simplest ones are given by a linear PDE which yields an algebraic relation between jet coordinates. This is what will be developed in the next section.

\sect{The Forsyth frames} \label{FL}

Over a Riemann surface $\Sigma$, let us introduce the algebraic PDE of fixed
order $s$ with smooth coefficients and defined by  
\begin{eqnarray}
 L_s f \zbz = 0,\qquad \mbox{with }
 L_s=\sum_{j=0}^s a^{(s)}_{(s-j)}\zbz \prt_{(j)},\qquad
\mbox{where  }a^{(s)}_{(0)}\zbz=1, \mbox{ and } a^{(s)}_{(1)}\zbz=0, 
\lbl{lr} 
\end{eqnarray}
When $\ovl{z}$ is viewed to play the role of a parameter 
the PDE is considered as an ODE in the independent variable $z$ and
the function $f$ is the unknown. It thus introduces a chiral splitting
between the complex coordinates.
Around any point of $\Sigma$ this ODE admits $s$ linearly independent
{\em local} solutions $f^{(R)}$, 
$R=1,\dots,s$ on a small enough neighborhood of any point.
Actually, any solution turns out to be a scalar density 
under holomorphic changes of charts $(U,z)\rightarrow (\widehat{U},w(z))$ with 
conformal weight $\frac{1-s}{2}$ in order to have a well
defined covariance on the Riemann surface $\Sigma$ which yields 
\begin{eqnarray}
L_s \wbw = {(w')}^{-\sm{1 + s}{2}}L_s\zbz \, {(w')}^{\sm{1 -
    s}{2}}\qquad \mbox{on } U\cap \widehat{U} \neq\emptyset.
\lbl{boltransform}
\end{eqnarray}
Recall that $L_sf$ has conformal weight $\frac{1+s}{2}$. For an overview see e.g. \cite{Gieres:1993} and references therein. Equation \rf{lr} can be recast as a first-order differential operator if the jet of order $s-1$ of any solution $f$ is considered as the variable. One has
\begin{eqnarray}
\biggl(\prt+A^{(s)}\zbz\biggr)\left(\begin{array}{c}
f\zbz\\  \prt f\zbz\\ \vdots\\ \prt_{(r-1)}f\zbz
\end{array} \right) = \left(\begin{array}{c} 
0 \\ \vdots \\ 0 \\ L_s f \zbz \end{array} \right) = 0 
\lbl{solutionsA}
\end{eqnarray}
where the $s\times s$ matrix 
\begin{eqnarray}
A^{(s)}\zbz = \left(
\begin{array}{ccccc} 
0 & - 1 & 0 & \dots & 0\\ 
0 & 0 & - 1 & \ddots & \vdots\\
\vdots & \vdots & \ddots & \ddots & 0 \\ 
0 & 0 & \dots & 0 & -1 \\
a^{(s)}_{(s)}\zbz & a^{(s)}_{(s-1)}\zbz & \dots & a^{(s)}_{(2)}\zbz & a^{(s)}_{(1)}\zbz 
\end{array}\right)
\lbl{As}
\end{eqnarray}
has entries
\begin{eqnarray}
[A^{(s)}\zbz]_{lm} = \begin{cases}
-1 & \mbox{for } m = l+1, \\
a^{(s)}_{(s-m+1)}\zbz & \mbox{for } l=s, \mbox{ with } a^{(s)}_{(1)}\zbz \equiv 0,\\
0 & \mbox{otherwise.}
\end{cases}
\end{eqnarray}
Moreover, each of them in the last row carries a $z$ lower index
content of covariant order $s-m+1$. But in account of \rf{solutionsA},
$A^{(s)}$ is expected to carry
a covariant index $z$ as the derivative $\prt_z$.

This allows one to associate to the ODE a system of $s$ identical
equations by introducing $s$ unknowns $f^{(R)}$, $R=1,\dots,s$.  So
it is a system of the same ODE over $\Sigma\times\mathbb{C}^s$.
Note that any linear change in the local linearly independent
solutions $\widetilde{f}^{(R)}\zbz 
= A^{(R)}_{(S)}
f^{(S)}\zbz$ over $U\subset\Sigma$ preserves \rf{lr}, $A\in
GL(s,\mathbb{C})$. For the time being, the matrix $A$ does not depend
on the local complex coordinates $\zbz$
on $\Sigma$, but this issue should be tackled as the gauging of
the largest symmetry group of the ODE \rf{lr}.

However, adapting Gunning \cite{Gun68} for a
well defined ODE over $\Sigma$ according to \rf{boltransform} if one
selects $s$ linearly independent local solutions of \rf{lr}, on the one hand,
$f^{(R)}\zbz$ on the coordinate neighborhood $U\subset\Sigma$, and on
the other hand, $\widehat{f}^{(R)}\wbw$ on the coordinate neighborhood
$\widehat{U}\subset\Sigma$, then on the overlapping of these
coordinate neighborhoods one has, in full generality, the following
gluing rule for solutions
\begin{eqnarray}
  \mbox{on } U\cap \widehat{U} \neq\emptyset, \qquad
\widehat{f}^{(R)}\wbw = (w')^{-\frac{1-s}{2}}\, T^{(R)}_{(S)} f^{(S)}\zbz
\lbl{f_glueing}
\end{eqnarray}
where the unique constant matrix $T$ tuns out to be a 1-cocycle of the
chosen coordinate covering on $\Sigma$ with values in
$GL(s,\mathbb{C})$. To prove \rf{f_glueing}, for the given
linearly independent $s$ solutions
$\widehat{f}^{(R)}\wbw$ in the open set
$\widehat{U} \subset\Sigma$, let us introduce the
functions $g^{(R)}\zbz$ on $U\cap \widehat{U} \subset U$ defined by
\begin{eqnarray}
  g^{(R)}\zbz = (w')^{\frac{1-s}{2}} \widehat{f}^{(R)}\wbw\qquad
 \mbox{on } U\cap \widehat{U} \neq\emptyset.
\lbl{def_g}
\end{eqnarray}
Upon using the covariance law \rf{boltransform} it is readily seen that
the $s$ functions
$g^{(R)}\zbz$ are linearly independent solutions of \rf{lr} over
$U\cap \widehat{U}\subset U$. Accordingly, for any other $s$ linearly
independent $f^{(R)}\zbz$ of \rf{lr} in the open
set $U$, the functions $g^{(R)}\zbz$ are unique linear combinations
of the functions $f^{(R)}\zbz$, that is $g^{(R)}\zbz =
T^{(R)}_{(S)}\,f^{(S)}\zbz$ a fact which demonstrates \rf{f_glueing}.

According to \cite{Forsyth59}, (see also \cite{Gov95}) one may define
{\em locally} $s-1$ smooth scalar 
fields over a neighborhood of any point of $\Sigma$
as quotients of $s-1$  local solutions by a preferred one which does
not vanish on a neighborhood of a given point, say on $U$,
\begin{eqnarray}
Z^{(R)}\zbz = \frac{f^{(R+1)}\zbz}{f^{(1)}\zbz},\qquad
R = 1,\cdots, s-1
\lbl{ZR}
\end{eqnarray}
where the functions $f^{(R+1)}$ and $f^{(1)}$ belong to
the same set of linearly independent solutions over $U$.
By virtue of \rf{f_glueing} one checks that
\begin{eqnarray}
  \mbox{on } U\cap \widehat{U} \neq\emptyset, \qquad
  \widehat{Z}^{(R)}\wbw = \frac{T^{(R+1)}_{(S+1)} Z^{(S)}\zbz +
  T^{(R+1)}_{(1)}}{T^{(1)}_{(S+1)} Z^{(S)}\zbz + T^{(1)}_{(1)}},
\lbl{Z_glueing}
\end{eqnarray}
a transformation law which shows that the local scalar fields $Z$ have to be
transformed in a homographic way in accordance with the Zucchini's
point of view \cite{Zuc93} on $\cW$-algebras.

Moreover, note that any linear change in the linearly independent
solutions $f^{(R)}$ on $U$ induces a
homographic transformation in the $Z^{(R)}$ on $U$ as
\begin{eqnarray}
  \widetilde{Z}^{(R)}\zbz = \frac{A^{(R+1)}_{(S+1)} Z^{(S)}\zbz +
  A^{(R+1)}_{(1)}}{A^{(1)}_{(S+1)} Z^{(S)}\zbz + A^{(1)}_{(1)}}.
\lbl{Z_gauge}
\end{eqnarray}
The question of gauging whether or not the matrix $A\in GL(s,\mathbb{C})$ into
$A\zbz$, in other words, render local the above transformation law
should be also tackled in the sequel.

This is the point where now one has to decide if $Z^{(R)}$ is a genuine
  scalar field on $\Sigma$ or not, namely keeping $T$ as general as
  possible or reduce it to the identity. If one chooses the latter
  what would be the meaning of $T=I$ for the space of solutions of the
  ODE \rf{lr}~? A possible answer would be that there exists basis of
  solutions which is globally defined on $\Sigma$ according to
  \rf{f_glueing} with $T=I$.
But there is a more precise statement which is the following. Consider
\begin{eqnarray}
\mbox{on } \widehat{U},\qquad
\widehat{Z}^{(R)}\wbw =
\frac{\widehat{f}^{(R+1)}\wbw}{\widehat{f}^{(1)}\wbw},\qquad
R = 1,\cdots, s-1
\end{eqnarray}
and thus with \rf{def_g} one can define on the overlapping the scalar functions
\begin{eqnarray}
   \mbox{on } U\cap \widehat{U} \neq\emptyset, \qquad
 \zeta^{(R)}\zbz = \frac{g^{(R+1)}\zbz}{g^{(1)}\zbz} =
 \widehat{Z}^{(R)}\wbw,\qquad R = 1,\cdots, s-1
\end{eqnarray}
which thus coincide with the scalar functions $\widehat{Z}^{(R)}$ on
the intersection $U\cap \widehat{U} \neq\emptyset$. Since in the glueing
rule given by \rf{f_glueing} the matrix $T$ only depends on the two
coordinate neighborhoods $U$ and $\widehat{U}$, let us perform the
following linear change $\widetilde{f} = a\,T f$ of linearly independent
solutions on the open set $U$, with given $a:\ U\rightarrow
\mathbb{C}^*$. This yields a gauge transformation over $U$ of the type
\rf{Z_gauge} with $A = T$ so that
\begin{eqnarray}
   \mbox{on } U\cap \widehat{U} \neq\emptyset, \qquad
 \zeta^{(R)}\zbz = \widehat{Z}^{(R)}\wbw = \widetilde{Z}^{(R)}\zbz,
\qquad R = 1,\cdots, s-1 .
\lbl{Z_scalar}
\end{eqnarray}
Hence by these redefinitions through gauge transformations, one can construct
$s-1$ scalar fields on $\Sigma$, still denoted by $Z^{(R)}$, each of
those being a collection of scalar maps defined on the various coordinate
neighborhoods satisfying \rf{Z_scalar} as matching rule.

Suppose now that a family of linear differential equations of the type \rf{lr} is indexed by the order $r\geq 2$. Accordingly, solutions must be labeled by the order $r$, and the above construction holds for each of the orders $r$. One may state the

\begin{Conjecture}
Around each point of the Riemann surface $\Sigma$ and for each integer
$r\geq2$, there is a map, $\Sigma \rightarrow \mathbb{C}P^{r-1}$ which
defines local inhomogeneous coordinates on $\mathbb{C}P^{r-1}$,
collectively denoted by the vector valued in $\mathbb{C}^{r-1}$ smooth
function on $\Sigma$ 
\begin{eqnarray}
  \vec{Z}\zbzr = (Z^{(1)}\zbzr,\dots,Z^{(r-1)}\zbzr),
\end{eqnarray}
where the $r-1$ components are given by
\begin{eqnarray}
Z^{(R)}\zbzr &=& \frac{f^{(R+1)}\zbzr}{f^{(1)}\zbzr},\qquad R=1,\cdots,r-1.
\lbl{ZR1}
\end{eqnarray}
$\vec{Z}\zbzr$ will be called a Forsyth frame.
For a given a point on $\Sigma$, all the frames must be equivalent for all
the physical points of view.
\end{Conjecture}

Returning to the general discussion with a given equation \rf{lr} of
order $s$ which introduces the truncation in the jet order, the
following theorem comes as a by-product. 
\begin{Statement} \label{Stamnt1}
For any $m\geq 1$, given an order $s$ of truncation dictated by a
differential equation of the type \rf{lr}, one has
\begin{eqnarray}
\prt_{(m)} \vec{Z}\zbz  =\sum_{l=1}^{s-1}
\cR_{(m)}^{(l)}\zbz\prt_{(l)}\vec{Z}\zbz,\quad \mbox{and}\quad
\cR_{(m)}^{(l)}\zbz=\delta_{(m)}^{(l)}\quad \mbox{if } 1\leq m \leq
s-1
\lbl{pdes}
\end{eqnarray}
The decomposition is universal for all the inhomogeneous 
coordinates, $Z^{(R)}\zbz$ in the sense that each $\cR_{(m)}^{(l)}$
does not depend on the index $(R)$ of the former. The $\cR$'s
correspond exactly to those heuristically introduced in \rf{Z_trunca}
and are specific to the order $s$ of truncation imposed by \rf{lr}. However the 
vectorial character of $\vec{Z}$ gives rise to some additional compatibility
conditions between themselves.
\label{guess}
\end{Statement}
\begin{Proof}
The proof of Theorem \ref{guess} is trivial by direct computation.
Indeed we can compute $\prt_s f^{(P+1)}\zbz$ for all $P=1,\dots, s-1$ in
two different ways. The first one comes from the very definition
Eq.\rf{ZR} and the Leibniz rule 
\begin{eqnarray}
\prt_{(s)} f^{(P+1)} = \prt_{(s)} \Big(Z^{(P)} f^{(1)}
\Big) = \prt_{(s)} Z^{(P)} f^{(1)}
+ Z^{(P)} \prt_{(s)} f^{(1)} + \sum_{j=1}^{s-1} \binom{s}{j}
\prt_{(j)} Z^{(P)} \prt_{(s-j)} f^{(1)}
\lbl{prtsfp1}
\end{eqnarray}
The second one comes from from the very definition
Eq.\rf{ZR} and the fact that both $f^{(P+1)}$ and $f^{(1)}$ are
solutions of Eq.\rf{lr}:  
\begin{eqnarray}
\prt_{(s)} f^{(P+1)} &=& 
- \sum_{j=0}^{s-1} a_{(s-j)}^{(s)} \prt_{(j)} \Big(Z^{(P)} f^{(1)}
\Big) 
= - \sum_{j=0}^{s-1} a_{(s-j)}^{(s)} \sum_{m=0}^j \binom{j}{m}
\prt_{(m)} Z^{(P)} \prt_{(j-m)} f^{(1)} \nn\\
&=& -\, Z^{(P)} \sum_{j=0}^{s-1} a_{(s-j)}^{(s)}
\prt_{(j)} f^{(1)}\zbz - \sum_{j=1}^{s-1} a_{(s-j)}^{(s)} 
\sum_{m=1}^j \binom{j}{m}
\prt_{(m)} Z^{(P)} \prt_{(j-m)} f^{(1)} \nn\\
&=& Z^{(P)} \prt_{(s)} f^{(1)} - \sum_{m=1}^{s-1} \prt_{(m)} Z^{(P)}
\sum_{j=m}^{s-1} a_{(s-j)}^{(s)} \binom{j}{m}
\prt_{(j-m)} f^{(1)} 
 \lbl{prtsfp2}
\end{eqnarray}
A direct comparison between Eqs.\rf{prtsfp1} and \rf{prtsfp2} entails
\begin{eqnarray}
\prt_{(s)} Z^{(P)} = \frac{-1}{f^{(1)}} 
\sum_{m=1}^{s-1} \left( \binom{s}{m} \prt_{(s-m)} f^{(1)} 
+ \sum_{j=m}^{s-1} a_{(s-j)}^{(s)} \binom{j}{m}
\prt_{(j-m)} f^{(1)} \right) \prt_{(m)} Z^{(P)}
\end{eqnarray}
The decomposition \rf{pdes} combined with the non vanishing of the
Wronskian determinant \rf{wronskian} yields 
\begin{eqnarray}
\cR_{(s)}^{(m)}\zbz \equiv
\frac{-1}{f^{(1)}\zbz} 
\left( \binom{s}{m} \prt_{(s-m)} f^{(1)}\zbz
+ \sum_{j=m}^{s-2} a_{(s-j)}^{(s)}\zbz \binom{j}{m}
\prt_{(j-m)} f^{(1)}\zbz \right)
\lbl{Rs}
\end{eqnarray}
where $\cR_{(s)}^{(m)}$ for $m=1,\dots,s-1$ depend on the coefficients
$a_{(s-j)}^{(s)}$ and $f^{(1)}$ and its $z$ derivatives up to order
$s-2$ since $a_{(1)}^{(s)} = 0$.
It is readily seen that the decomposition does not depend on the index
of the solution $f^{(P+1)}$.
One can extend \rf{Rs} to the case $m=0$ since $f^{(1)}$ is solution
of \rf{lr} by setting $\cR_{(s)}^{(0)} \equiv 0$.
\end{Proof}
Let us introduce the Wronskian as a $(s-1)\times(s-1)$-matrix
\begin{eqnarray}
\varpi\zbz = \left(\varpi^{(R)}_{(\ell)}\zbz \right) = 
\begin{pmatrix}
\prt Z^{(1)}\zbz&\cdots&\prt Z^{(s-1)}\zbz\\
\vdots &\ddots& \vdots \\
\prt_{(s-1)} Z^{(1)}\zbz&\cdots&\prt_{(s-1)} Z^{(s-1)}\zbz\\
\end{pmatrix} .
\lbl{wronskian}
\end{eqnarray}
Hence in the algebra of squared
matrices of order $s-1$ the relationships \rf{pdes} state that any
$z$-derivative of the 
Wronskian $\varpi$ can be decomposed as a product of a rectangular
matrix with the functions $\cR$ as entries by $\varpi$, in details, 
\begin{eqnarray}
  \prt_{(m)} \varpi^{(R)}_{(\ell)} = \sum_{k=1}^{s-1}
  \cR_{(m+\ell)}^{(k)}\varpi^{(R)}_{(k)} .
  \lbl{DOR}
\end{eqnarray}
In order to be the most general as possible, the Wronskian may be extended to a $m\times(s-1)$ rectangular matrix when higher derivatives $m\geq s$ of the $Z^{(R)}$'s are considered. In account of \rf{pdes}, the rectangular matrix of derivatives of $Z$ up to order $m$ can always be expressed in terms of the Wronskian matrix \rf{wronskian}.
We shall call this mechanism connected to the truncation heuristically introduced in
\rf{bracket} as a Derivative Order
Reduction, or in shorthand D.O.R.
Note also that thanks to \rf{pdes} a straightforward computation gives
\begin{eqnarray}
\cR_{(s)}^{(s-1)}\zbz = \prt\ln\det\varpi\zbz.
\lbl{Rs1}
\end{eqnarray}
The preferred solution $f^{(1)}$ which crucially takes place in the
computation of the $\cR$'s plays a distinguished role in the
construction as it has been already seen.
In particular it infers a linear relationship for
$f^{(1)}$ with $j=s-1$ in Eq.\rf{Rs}
\begin{eqnarray}
\cR_{(s)}^{(s-1)}\zbz=-s\,\prt \ln f^{(1)}\zbz
\end{eqnarray}
which yields, on the one hand, together with \rf{Rs1}
\begin{eqnarray}
        f^{(1)}\zbz = (\det \varpi \zbz)^{-1/s},
        \lbl{f1}
\end{eqnarray}
and on the other hand,
\begin{eqnarray}
\Longrightarrow \prt f^{(1)} = \cQ_{(1)} f^{(1)}, \qquad
\mbox{where } \cQ_{(1)} =-\sm{1}{s}\cR_{(s)}^{(s-1)}
\lbl{Rss-1}
\end{eqnarray}
and by successive $z$-derivatives one gets a recursive formula
\begin{eqnarray}
\prt_{(n)}f^{(1)}\zbz=\cQ_{(n)}\zbz f^{(1)}\zbz, \quad \mbox{with }
\cQ_{(n)} = \prt\cQ_{(n-1)} + \cQ_{(n-1)} \cQ_{(1)},\quad \mbox{and
}\cQ_{(0)} = 1, 
\lbl{recurf1}
\end{eqnarray}
so that $\cQ_{(n)}$ turns to be a differential polynomial in
$\cQ_{(1)}$ (i.e. in $\cR_{(s)}^{(s-1)}$), namely $\cQ_{(n)} = (\prt +
\cQ_{(1)} )^{n-1} \cQ_{(1)}$. Using \rf{recurf1} into \rf{Rs} and 
eliminating $f^{(1)}$ allow to write a linear system in a Gauss form
with respect to the $a$'s coefficients
\begin{eqnarray}
  \cR_{(s)}^{(j)} + \begin{pmatrix} s \\ j \\ \end{pmatrix} \cQ_{(s-j)}
+ \sum_{l=j}^{s-1} 
\begin{pmatrix} \ell \\ j \\ \end{pmatrix} a_{(s-\ell)}^{(s)}
\cQ_{(\ell-j)} = 0,\qquad \mbox{ for } j = 0,\dots,s-1,
\lbl{aQR}
\end{eqnarray}
which expresses the relationship between the $a^{(s)}$'s and the
$\cR_{(s)}$'s. This step is independent of $f^{(1)}$ provided that the
$\cR_{(s)}$'s 
are given (together with some compatibility conditions) and we will
consider from now on and throughout the all paper that the degrees of
freedom will be the $\cR_{(s)}$'s.
Hence, solving iteratively the system \rf{aQR} with respect to the
$a^{(s)}$'s one gets 
\begin{eqnarray}
  \begin{cases}
a_{(1)}^{(s)} = 0 & \mbox{for } j = s-1 \\
{\dps a_{(2)}^{(s)} = -\cR_{(s)}^{(s-2)} - \binom{s}{s-2} 
\cQ_{(2)} } & \mbox{for } j = s-2 \\
{\dps a_{(3)}^{(s)} = -\cR_{(s)}^{(s-3)} - \binom{s}{s-3}
\cQ_{(3)} - \binom{s-2}{s-3}
a_{(2)}^{(s)} \cQ_{(1)}} & \mbox{for } j = s-3 \\
\qquad\vdots & \mbox{and so on up to } j = 0.
\end{cases}
\end{eqnarray}
This shows that to a given a DOR \rf{pdes} there corresponds a
holomorphically covariant differential equation of the type \rf{lr}
whose smooth coefficients are expressed as differential polynomials in
the structure function $\cR_{(s)}^{(s-1)}$ and linearly with respect
to the others.

Moreover one has the following property which is exactly the
compatibility condition \rf{prtRJacobi}. 
\begin{Properties}
For $p=1,\dots,s-1$ and $n\geq s$
\begin{eqnarray}
\cR^{(p)}_{(m+n)}\zbz = \sum_{j=0}^{m}\left(\begin{array}{c}m\\j
\end{array}\right)
\sum_{\ell=1}^{s-1}
\prt_{(j)}\cR^{(\ell)}_{(n)}\zbz\cR_{(m+\ell-j)}^{(p)}\zbz
\lbl{prop}
\end{eqnarray}
where $1\leq l\leq s-1$; so, recursively all the
$\cR^{(l)}_{(s+m)}\zbz\quad m>0$ coefficients can be derived from
the basic $\cR^{(l)}_{(s)}\zbz$ ones.
\end{Properties}
In particular, for the case $m=1$, one has
\begin{eqnarray}
  \cR^{(p)}_{(n+1)}\zbz = \prt\cR^{(p)}_{(n)}\zbz + \sum_{\ell = 1}^{s-1}
\cR^{(\ell)}_{(n)}\zbz \cR_{(\ell + 1)}^{(p)}\zbz,
\lbl{prop_1}
\end{eqnarray}
an equation which will be useful for future applications.

The basic $\cR^{(l)}_{(s)}$'s, namely the structure functions given in
the introductory section, play a central role and it would be
worthwhile to have some hints about their geometric nature. In order
to be closer as possible to a differential geometric setting for our
approach, let us proceed as follows.  For the the $z$-jet of a fixed
order, one has the following holomorphic glueing rules under the
change $z\mapsto w = w(z)$
\begin{eqnarray}
 \prtz^{k+1}Z\zbz = 
\begin{cases}
w'(z)\,\prt_w Z\wbw & \qquad \text{if } k=0,\\[2mm]
w^{(k+1)}(z) \prt_w Z\wbw + {\dps 
\sum_{\ell=1}^k \prt_w^{\ell + 1} Z\wbw \times}& \\[4mm] 
{\dps \quad\qquad \times
\sum_{r=\ell}^k
 \frac{k!}{(k-r)!}\,w^{(k-r+1)}(z)  \left( 
\sum_{\scriptstyle a_1+\cdots+ r a_r = r \atop \scriptstyle
  a_1+\cdots+  a_r = \ell} 
\left(\prod_{n=1}^r \frac{1}{a_n!}
\left(\frac{w^{(n)}(z)}{n!}\right)^{a_n} \right) \right) } &\qquad
  \text{if } k\geq 1,
\end{cases}
\lbl{jetglueing}
\end{eqnarray}
where the last expression comes from the use of the Fa\`a di Bruno
formula for higher order chain rule of derivatives. For further
calculations, one has more explicitly and under a more elegant form, 
for $k\geq 3$
\begin{eqnarray}
  (\prt_{(k+1)} Z)\zbz &=& (w')^{k+1} (\prt_{(k+1)} Z)\wbw 
+ \binom{k+1}{2} (w')^{k} \prtz\ln w'\, (\prt_{(k)} Z)\wbw\nn\\
&& +\ \binom{k+1}{3} (w')^{k-1}
  \Big( \{w,z\} + \sm{3}{4}\,k (\prtz\ln w')^2 \Big) (\prt_{(k-1)}
  Z)\wbw \lbl{forcalc} \\
&& +\ \binom{k+1}{4} (w')^{k-2} \Big( \prtz \{w,z\} +
  2(k-1) \{w,z\}\prtz\ln w' 
  + \binom{k}{2} (\prtz\ln w')^3 \Big) (\prt_{(k-2)} Z)\wbw \nn\\
&& +\ \mbox{lower order derivatives}, \nn
\end{eqnarray}
where $\lbrace w,z\rbrace = \prtz^2 \ln
w' - \sm{1}{2}(\prtz\ln w')^2 = \frac{w'''}{w'} -
\sm{3}{2}\left(\frac{w''}{w'}\right)^2$ denotes the Schwarzian
derivative. 

Now for fixed $s$, one can obtain the geometric properties 
of the $s-1$ structure functions $\cR$'s by solving the linear system
\rf{pdes} with respect to the $\cR$'s by Cramer
method, one gets the following Lie form associated to the PDEs
\rf{pdes} (see eg \cite{Pom3}) for $m=1,\dots,s-1$
\begin{eqnarray}
\Phi^{(m)}(\vec{Z}_s):= (-1)^{s-1-m}\
  \frac{\det(\prt\vec{Z},\prt_{(2)}\vec{Z},\dots,\widehat{\prt_{(m)}\vec{Z}},
  \dots,\prt_{(s-1)}\vec{Z},\prt_{(s)}\vec{Z})}{\det\varpi} 
= \cR_{(s)}^{(m)}\zbz ,
\lbl{diff_inv}
\end{eqnarray}
where the symbol $\widehat{\quad}$ means omission. This expression can
simply be rewritten as 
\begin{eqnarray}
  \cR_{(s)}^{(m)} = \prt_{(s)} Z^{(R)} [\varpi^{-1}]_{(R)}^{(m)}.
\end{eqnarray}
For $s=2$, one has the obvious relations $\cR_{(m)}^{(1)} = \prt_{(m)}
Z / \prt_{(1)} Z$. 
According to an approach advocated by Vessiot to the
Picard-Vessiot theory \cite{Pom3} (and references
therein) one can construct, regarding the 
present case and by a repeated use of \rf{forcalc}, a natural
holomorphic bundle with a $(s-1)$-dimensional 
fiber with fiber coordinates $(u^{(1)},\dots,u^{(s-1)})$. It is defined by
the following holomorphic transition functions induced by the holomorphic 
change of chart $w = \varphi(z)$ on $\Sigma$
\begin{eqnarray}
\begin{cases}
  w = \varphi(z) \\[2mm]
U^{(s-1)} = \frac{1}{w'}\Big( u^{(s-1)} - 
\binom{s}{2} \prt\ln
w'\Big) \qquad\quad \mbox{affine bundle !} \\[2mm]
U^{(s-2)} = \frac{1}{w'^2}\Big( u^{(s-2)} + 
\binom{s-1}{2} u^{(s-1)} \prt\ln w' - 
\binom{s}{3} \big( \{w,z\}
+ \sm{3}{4}\binom{s-1}{1} (\prt\ln w')^2 \big) \Big) \\[2mm]
U^{(s-3)} = \frac{1}{w'^3}\Big( u^{(s-3)} + \binom{s-2}{2} u^{(s-2)}\prt\ln w'
+ u^{(s-1)} \binom{s-1}{3} \big( \{w,z\}
+ \sm{3}{4}\binom{s-2}{1} (\prt\ln w')^2 \big) \\
\hskip 2.5cm - \binom{s}{4} \big( \prt \{w,z\} + 2\binom{s-2}{1} \{w,z\}\prt\ln w' + \binom{s-1}{2} (\prt\ln w')^3 \big) \Big) \\
\quad \vdots \\
U^{(1)} = \mbox{a very intricate expression depending on all the } u^{(i)}\mbox{'s}
\end{cases}
\lbl{bdle}
\end{eqnarray}
where the transition laws become more and more involved. This bundle can be recast into a holomorphic natural bundle of
geometric objects 
(but however with smooth sections $\cR$ in accordance with locality) as
fibered product over the Riemann surface $\Sigma$
$${\cal F}_{\mathrm{affine}}\times_{\Sigma}\ {\cal F},$$
where ${\cal F}_{\mathrm{affine}}$ is the affine bundle and ${\cal F}$
is the bundle with very intricate remaining but  important patching
rules for the sequel.
Having at our disposal some of the main glueing rules of the fundamental $\cR$'s, it is possible to obtain the geometrical nature of some of the coefficients of \rf{lr}. Indeed, in terms of the Wronskian $\cR_{(s)}^{(s-1)} = \prt \ln\det\varpi$, one
finds for the coefficient
\begin{eqnarray}
  a_{(2)}^{(s)} = \sm{s-1}{2} \left(\prt \cR_{(s)}^{(s-1)} -
  \sm{1}{s} (\cR_{(s)}^{(s-1)})^2 \right) - \cR_{(s)}^{(s-2)},
\lbl{projconnec}
\end{eqnarray}
while both $\cR_{(s)}^{(s-1)}$ and $\cR_{(s)}^{(s-2)}$ glue as
smooth sections of the bundle defined by \rf{bdle}.  After a direct
computation 
\begin{eqnarray}
  a_{(2)}^{(s)} \zbz = (w')^2 a_{(2)}^{(s)}\wbw +
  \sm{s(s^2-1)}{12} \left( \prt^2 \ln w' - \sm{1}{2}(\prt\ln w')^2 \right)
\end{eqnarray}
which shows that $a_{(2)}^{(s)}$ is proportional to a projective connection
as is well known, since the inhomogeneous term in the glueing rule is
nothing but the Schwarzian derivative $\lbrace w,z\rbrace = \prt^2 \ln
w' - \sm{1}{2}(\prt\ln w')^2 $. The projective connection is constructed
over the frame $\vec{Z}$ according to \rf{projconnec}.

\begin{Remark}
For the case $s=3$, one has $\cR^{(2)}_{(3)} = \prt \ln\det\varpi$, and
$\cR^{(1)}_{(3)} = \det(\prt^3\vec{Z},\prt^2\vec{Z})/\det\varpi$. With
$\cQ_{(1)} = \sm{-1}{3} \cR^{(2)}_{(3)}$, one readily gets 
\begin{eqnarray}
a^{(3)}_{(2)} &=& - \cR^{(1)}_{(3)} - 3 \cQ_{(2)} = -
  \cR^{(1)}_{(3)} - 3 (\prt \cQ_{(1)} + (\cQ_{(1)})^2) \nn\\
&&\lbl{case_3} \\
a^{(3)}_{(3)} &=& \cR^{(1)}_{(3)} \cQ_{(1)} + 2 (\cQ_{(1)})^3
  -\prt^2 \cQ_{(1)} = \sm{1}{3}\left(\prt a^{(3)}_{(2)} + \prt
  \cR^{(1)}_{(3)} + 
  \sm{2}{3} \cR^{(2)}_{(3)} a^{(3)}_{(2)} 
- \sm{1}{3}\cR^{(1)}_{(3)} \cR^{(2)}_{(3)} \right) \nn
\end{eqnarray}
which are exactly those coefficients obtained for the so-called $\cW_3$-algebra
\cite{Forsyth59,BaLa01}. The last expression is given in
terms of the projective connection and the structure functions only.
\label{rem_s=3}
\end{Remark}

\begin{Remark}
The factorization property of the differential operator $L_s$ of order
$s$ in terms of 1st order differential operators with nowhere vanishing
coefficients can be obtained if
and only if $L_s$ is a non-oscillating operator, see \cite{GO94} for
some details. For more concreteness, let us illustrate this
factorization property for $s=2,3$.
\begin{enumerate}
\item The $s=2$ case. Take $f^{(1)}=(\prt Z)^{-1/2}=:\lam^{-1/2}$ as a
  nowhere vanishing particular solution of $L_2 f = 0$. One can write
  \begin{eqnarray}
    L_2 = \prt^2 + a^{(2)}_{(2)} = (\prt - b)(\prt + b)\quad
    \mbox{with } a^{(2)}_{(2)} = \prt b - b^2,\quad \mbox{and } b = 
    -\prt\ln f^{(1)} =: - \cQ_{(1)},
  \end{eqnarray}
One finds the expected factorization
\begin{eqnarray}
L_2 = \big(\prt - \sm{1}{2}\prt\ln\prt\lam\big) \big(\prt +
\sm{1}{2}\prt\ln\prt\lam\big),  
\qquad \mbox{and }\quad a^{(2)}_{(2)} = \sm{1}{2} \prt^2\ln\prt\lam -
\sm{1}{4}(\prt\ln\prt\lam)^2.
\end{eqnarray}
\item The $s=3$ case amounts to writing
  \begin{eqnarray}
    L_3 = (\prt - b_1 - b_2) (\prt + b_2) (\prt + b_1)
  \end{eqnarray}
with $b_1 = -\prt\ln f^{(1)}= - \cQ_{(1)}$ and a possible choice for
$b_2$ is given by $b_2 = b_1 - \prt\ln\prt Z^{(1)}$ --it could be
possible to choose $b_2 = b_1 - \prt\ln\prt Z^{(2)}$ since $b_1$ never
vanishes. Then substituting into
\begin{eqnarray*}
  a^{(3)}_{(2)} &=& \prt(b_1+b_2) - (b_1 + b_2)^2 + \prt b_1 + b_1 b_2
  \\
a^{(3)}_{(3)} &=& \prt^2 b_1 + \prt(b_1 b_2) - (b_1+b_2) (\prt b_1 + b_1 b_2),
\end{eqnarray*}
one exactly recovers the expressions given above in \rf{case_3} for
the coefficients of $L_3$.
\end{enumerate}
\end{Remark}
Still with a fixed given order $s$, it is possible to construct a
connection-like object. 
With the Dolbeault decomposition of the de Rham differential $\mathrm{d}=
\boldsymbol{\prt} + \ovl{\boldsymbol{\prt}}$ let us define the flat connection
(pure gauge)
\begin{eqnarray}
  \cJ = \mathrm{d}\varpi\,\varpi^{-1} = 
\boldsymbol{\prt}\varpi\,\varpi^{-1} +
\ovl{\boldsymbol{\prt}}\varpi\,\varpi^{-1} = \mathrm{d}z \cJ_z +
\mathrm{d} \ovl{z}\cJ_{\ovl{z}}.
\lbl{J_connec}
\end{eqnarray}
Obviously its curvature vanishes
\begin{eqnarray}
 \cF = \mathrm{d}\cJ - \cJ^2 =  0\quad\Longrightarrow\quad 
\prtbz \cJ_z - \prtz\cJ_{\ovl{z}} + [\cJ_z ,\cJ_{\ovl{z}}] = 0.
\lbl{J_curv} 
\end{eqnarray}
The $(1,0)$-component of the $(s-1)\times(s-1)$-matrix connection of
$\cJ$ is by construction 
\begin{eqnarray}
{\cJ_{\z}}_{(m)}^{(n)}\zbz\equiv \sum_{R=1}^{s-1} \prt\varpi_{(m)}^{(R)}\zbz
[\varpi^{-1}]_{(R)}^{(n)}\zbz = \cR_{(m+1)}^{(n)}\zbz,\qquad
m,n = 1,\dots, s-1
\lbl{DS00}
\end{eqnarray}
or more explicitly in matrix form
\begin{eqnarray}
\cJ_{(z)}\zbz = \left(\begin{array}{cccccc}
0&1&0&0&\cdots&0\\
0&0&1&0&\cdots &\\
0&0&0&1&\cdots&0\\
\vdots&\vdots&\vdots&\vdots &\ddots& \vdots \\
0&0&0&0&\cdots&1\\
\cR_{(s)}^{(1)}\zbz&\cR_{(s)}^{(2)}\zbz&\cdots&\cdots&\cdots&
\cR_{(s)}^{(s-1)}\zbz  
\end{array}
\right). \lbl{JZ0}
\end{eqnarray}
This matrix turns out to be of Frobenius type (similar to the matrix
\rf{As}) and therefore $\cJ_z$ is not a Lie algebra-valued covariant
component of an usual connection. Note that the form of the matrix is
close to the Drinfeld-Sokolov one \cite{Drinfeld:1984qv}, but differs by the non-vanishing
term $\cR_{(s)}^{(s-1)}$.
Furthermore, it is useful to notice that 
\begin{eqnarray}
\cR_{(s)}^{(s-1)}\zbz= \mbox{Tr} \cJ_{(z)}\zbz = \mbox{Tr}(\prt\varpi\zbz\varpi^{-1}\zbz), 
\lbl{traccia}
\end{eqnarray}
in accordance with \rf{Rs1}.

\sect{B.R.S. approach} \label{BRS}

In this section, the heuristic presentation of the truncation
procedure and its consequences on the formulation of the algebra of
Ward operators given in Section \ref{Motiv} is translated into the BRS language.
As is well know, this allows to reformulate in more algebraic terms
the presentation of a symmetry, and in particular, will give a more
universal character of the possible variations on truncated Taylor
expansions of the scalar fields~$Z$.

\smallskip
 
Having still in mind that we are at a fixed given order $s$ for the
truncation \rf{Z_trunca} or \rf{pdes}, and by recalling Theorem \ref{Stamnt1}, we
can turn the $s-1$ local parameters $\Gamma^{(\ell)}$ to Faddeev-Popov
($\Phi\Pi$) ghosts $\cK^{(l)}$. The variation \rf{deltaZbis} can be recast
in a B.R.S. algebraic language as
\begin{eqnarray}
\delta_{\cWs} Z^{(R)}\zbzs = \sum_{\ell=1}^{s-1}\cK^{(\ell)}\zbzs \prt_{(\ell)}
Z^{(R)}\zbzs, \qquad
1\leq R \leq s-1,
\lbl{brsZ}
\end{eqnarray}
where the variation is given by a summand over the
independent derivatives  up to order $s-1$ due to the DOR of order
$s$.  The ghost fields $\cK^{(l)}$, of which number is  restricted to
the range $\ell = 1,\dots,s-1$, serve to define the  $\cWs$-algebra
relative to the truncation at the level $s$.  We emphasize that the
operation $\delta_{\cWs}$ which is required to be nilpotent,   depends
on the level $s$ of truncation. Accordingly, the ghost parameters
$\cK^{(l)}$ depend on the truncation process by their number, see
\rf{brsZ}, and  generate  a $\cW$-algebra once the level is fixed.  By
an argument based on the nilpotency, for $l=1,\dots,s-1$,
\begin{eqnarray}
\delta_{\cWs} \cK^{(l)}\zbzs
&=:& \sum_{m=1}^{s-1} \cK^{(m)}\zbzs \cB^{(l)}_{(m)}\zbzs \nn\\
&=& \sum_{m,n=1}^{s-1} \cK^{(m)}\zbzs 
\sum_{j=1}^m \binom{m}{j}
\prt_{(j)} \cK^{(n)}\zbzs \cR_{(n+m-j)}^{(l)}\zbzs,
\lbl{brsK}
\end{eqnarray}
where in the r.h.s the dependence in the level $s$ of truncation has been explicitly written. Note also that the products of two underived ghosts
drop out by $\Phi\Pi$ charge argument. This variation defines a $(s-1)\times(s-1)$-matrix $\cB$ carrying ghost number one, and thus depending on the level $s$ of truncation through the structure functions $\cR$ pertaining to that level $s$. In more details,
\begin{eqnarray}
        \cB^{(l)}_{(m)}\zbzs := \sum_{n=1}^{s-1} \sum_{j=0}^m \binom{m}{j}
\prt_{(m-j)} \cK^{(n)}\zbzs \cR_{(n+j)}^{(l)}\zbzs,
\lbl{B}
\end{eqnarray}
a remarkable combination over the ghosts $\cK^{(\ell)}$, $\ell = 1,\dots, s-1$ which could have been readily red of from the variation \rf{deltaR}. For the Wronskian \rf{wronskian} the $\cWs$-algebra extended to
\begin{eqnarray}
\delta_{\cWs} \varpi_{(\ell)}^{(R)}\zbzs
  = \sum_{n=1}^{s-1} \cB_{(\ell)}^{(n)}\zbzs \ \varpi_{(n)}^{(R)}\zbzs, 
\qquad \ell,R = 1,\dots,s-1
  \lbl{SderZ}
\end{eqnarray}
where the matrix product is understood for $(s-1)\times(s-1)$-matrices. 
One also have
\begin{eqnarray}
\delta_{\cWs} \varpi^{-1} = - \varpi^{-1} \cB ,
\lbl{sinvarpi}
\end{eqnarray}
and accordingly, using $\delta_\cW \mathrm{d} + \mathrm{d} \delta_\cW = 0$,
\begin{eqnarray}
        \delta_{\cWs} \cJ = - \mathrm{d}\cB + [\cB,\cJ],
        \lbl{deltaJ}
\end{eqnarray}
where the bracket is graded on forms with $(s-1)\times(s-1)$-matrix values .
The nilpotency property provides first,
 \begin{eqnarray}
 \delta_{\cWs} \cB\zbzs = \cB\zbzs\cB\zbzs = \cB^2\zbzs
\lbl{SB}
 \end{eqnarray}
 and second, by $\Phi\Pi$ argument
 \begin{eqnarray}
 \delta_{\cWs} \mbox{Tr} \Big(\cB\zbzs^{(2n+1)} \Big) = 0 ,\qquad n =0,1,\dots
\lbl{Soddtraces}
\end{eqnarray}
where Tr is the usual trace on matrices.
The B.R.S. variation of all the structure functions 
$\cR_{(n)}^{(p)}\zbzs$ ($p=1,\dots,s-1$ and even for $n\geq s$)  can
be directly found from the variation \rf{deltaR} to write
\begin{eqnarray}
\delta_{\cWs} \cR_{(n)}^{(p)}\zbzs
&=& \cB_{(n)}^{(p)}\zbzs -\sum_{q=1}^{s-1}\cR_{(n)}^{(q)}\zbzs
\cB_{(q)}^{(p)}\zbzs,
\lbl{sR} 
\end{eqnarray}
where $\cB$ defined above in \rf{B} may be extended to a rectangular
matrix for lower indices greater than $s-1$, while the upper ones are
still lower than this value imposed by the level of truncation, since
all the $\cR$'s can be gathered into a rectangular matrix. One checks
that it is compatible with the case
$\cR^{(k)}_{(\ell)}=\delta^{k}_{\ell}$ 
which is kept invariant.
We stress that while the first $\cB$ term of the r.h.s. of the
variation \rf{sR} is a rectangular matrix, the $\cB$ under the summand
is a squared one. As noted before the matrix $\varpi_{(m)}^R$ can be
taken to be a rectangular as well, when $m\geq s$, and the variation
\rf{SderZ} relative to the algebra $\cWs$ is accordingly modified by  
\begin{eqnarray}
\delta_{\cWs} \varpi_{(m)}^{(R)}\zbzs
  = \sum_{\ell=1}^{s-1} \cB_{(m)}^{(\ell)}\zbzs \ 
\varpi_{(\ell)}^{(R)}\zbzs, \qquad R = 1,\dots,s-1\mbox{ and } m \geq 1,
  \lbl{SderZmod}
\end{eqnarray}
where the matrix $\big(\cB_{(m)}^{(\ell)}\big)$ can be
rectangular. Supported by the fact 
that the ghosts $\cK^{(\ell)}$ are subordinated to the given level of
truncation $s$, one may now define
\cite{Ba88}, for $\ell =1, \dots, s-1$, the $\ell$-th derivatives
${\dps \prt_{(\ell)}=
  \bigg\{\frac{\prt}{\prt\cK^{(\ell)}},\delta_{\cWs} \bigg\} }$ as an
anticommutator, thus the DOR equation \rf{DOR} is recovered 
\begin{eqnarray}
        \prt_{(\ell)} \varpi_{(m)}^{(R)}\zbzs = \sum_{u=1}^{s-1} 
\cR^{(u)}_{(\ell+m)}\zbzs \varpi_{(u)}^{(R)}\zbzs, \qquad R,\ell=1,\dots,s-1,
 \mbox{ and } m\geq 1
\end{eqnarray}
This shows that the BRS algebra encapsulates the DOR mechanism just by
construction and provides a consistency of the present approach.

Furthermore, from Eq.\rf{SderZ} one has 
\begin{eqnarray}
\delta_{\cWs}\ln\det{\varpi\zbzs} = \mathrm{Tr}\,\cB\zbzs
\lbl{Sdetvarpi}
\end{eqnarray}
which gives for the variation of $\cR^{(s-1)}_{(s)}$,
\begin{eqnarray}
        \delta_{\cWs} \cR^{(s-1)}_{(s)}\zbzs =
 \prt\, \mathrm{Tr}\,\cB\zbzs = \mathrm{Tr}\,\prt\,\cB\zbzs.
\end{eqnarray}

\sect{Covariance under holomorphic reparametrization} \label{Covar}

The covariance property under holomorphic change of local coordinates on the
Riemann surface is analyzed  for the so far obtained quantities. As it
will be shown, this analysis will amount to switching to new ghost
fields of a tensorial nature in contrast to that of the $\cK$'s.
 
In view of the patching rules \rf{jetglueing} under finite holomorphic reparametrizations, it possibly renders more explicit some of the covariance properties of the theory relative to a fixed order $s$. As it will be shown, the study of covariance will appear as a key step in the construction of $\cW$-algebras.

Under finite holomorphic change of charts $z\lra w\z$ the covariance property of the $s-1$ scalar fields (emerging from the truncation at level $s$) writes
\begin{eqnarray}
Z^{(R)}\zbzs = Z^{(R)}\wbws, \qquad R = 1,\cdots, s-1
\end{eqnarray}
and implies that the Wronskian matrix $\varpi$ behaves as a non tensorial covariant quantity 
\begin{eqnarray}
\varpi_{(n)}^{(R)}\wbws = \sum_{m=1}^{s-1} \Phi_{(n)}^{(m)}\z \varpi_{(m)}^{(R)}\zbzs, 
\lbl{finiteWr}
\end{eqnarray}
and while for its inverse
\begin{eqnarray}
[\varpi^{-1}]^{(n)}_{(R)}\wbws
= \sum_{m=1}^{s-1} [\varpi^{-1}]_{(R)}^{(m)}\zbzs {\Phi^{-1}}_{(m)}^{(n)}\z
\end{eqnarray}
where patching rules are governed by the $(s-1)\times(s-1) $ lower
triangular holomorphic matrix $\Phi^{(\ell)}_{(k)}(z)$ 
for $k,\ell = 1, \dots, s-1$, depending on the Jacobian $w'$ and its
derivatives. Its inverse matrix is more easily computable and given by
(see~\rf{jetglueing})  
\begin{eqnarray}
[\Phi^{-1}]^{(\ell)}_{(k)}(z)=
 \left\{
 \begin{array}{lc}
w^{(k)}(z)\ , & \mbox{if } \ell=1 \\[2mm]
{\dps \sum_{r=\ell-1}^{k-1} \frac{(k-1)!\,w^{(k-r)}(z)}{(k-r-1)!}
\hskip -3mm \sum_{\tiny
\begin{array}{c}
{\tiny a_1+\cdots+ r a_r = r} \\
{\tiny a_1+\cdots+  a_r = \ell-1} \end{array} }\hskip -4mm
\left(\prod_{n=1}^r \frac{1}{a_n!}
\left(\frac{w^{(n)}(z)}{n!}\right)^{a_n}\right) , }
 & k\geq \ell \geq 2 \\
0, & k < \ell
\end{array}\right.
\lbl{phi}
\end{eqnarray}
with non vanishing  determinant, $\det \Phi^{-1}(z) =
\big(w'(z)\big)^{s(s-1)/2}$, since the diagonal entries are given by 
$[\Phi^{-1}]^{(\ell)}_{(\ell)}(z) = (w'(z))^\ell$. Note also that the
order of the matrix $\Phi^{(\ell)}_{(k)}(z)$ (or that of its inverse
as well) is subject to the order $s$ of truncation.  
Accordingly, since the variation \rf{brsZ} relative to the order $s$
has to behave as a scalar, the $s-1$ ghosts of the level $s$ of
truncation turn out to be contravariant quantities 
\begin{eqnarray}
\cK^{(l)}\wbws = \cK^{(m)}\zbzs[\Phi^{-1}]_{(m)}^{(l)}(z).
\lbl{changeK}
\end{eqnarray}
Taking into account of \rf{phi}, the ghosts $\cK^{(m)}\zbz$ behave as jets,
except for the top one of order $s-1$ which turns out to be a contravariant
 tensor of order $s-1$. 
For the sake of completeness, the behaviors 
 of the rectangular matrices $\cB^{(l)}_{(p)}\zbzs$  and $\cR^{(l)}_{(p)}\zbzs$ 
respectively come from \rf{finiteWr}, \rf{SderZ} and \rf{pdes} or \rf{sR}.
 They are respectively found to be
\begin{eqnarray}
\cB\wbws = \Phi\z \cB\zbzs \Phi^{-1}\z,
\lbl{changeB}
\end{eqnarray}
and for $\ell\leq s-1$,
\begin{eqnarray}
\cR_{(k)}^{(\ell)}\wbws = \sum_{m=1}^k \sum_{p=\ell}^{s-1}
 \Phi_{(k)}^{(m)}\z \cR_{(m)}^{(p)}\zbzs [\Phi^{-1}]_{(p)}^{(\ell)}\z.
\lbl{changeR}
\end{eqnarray}
Thanks to the their definition \rf{R[s]}, one obtains the following
identities for $k\leq s-1$, 
\begin{eqnarray}
[\Phi^{-1}]_{(k)}^{(k)}(z)\Phi_{(k)}^{(k)}\z = 1,\quad \mbox{(no
  summation)}, \qquad 
\sum_{u=\ell}^k [\Phi^{-1}]_{(k)}^{(u)}(z) \Phi_{(u)}^{(\ell)}\z = 0,
\quad \mbox{if } \ell< k\leq s-1. 
\lbl{Ids}
\end{eqnarray}

\sect{Jets versus tensors, or how to recover $\cW$-algebras} \label{W}

The algebra of transformations Eqs.\rf{brsZ} and \rf{brsK}  are written
in terms of ghosts  
which under holomorphic change of charts (see Eq.\rf{changeK}) behave
as jets, thus do not carry any tensorial nature. 

We want to show that these transformations encode a structure of
$\cW_s$-algebra if the D.O.R. 
mechanism is provided by a truncation at the $s$-th level. The latter
may be implemented by means of a given differential equation \rf{lr} which
serves to generate what it is called in the paper, the
 Forsyth frames. 
Since objects of jet nature are heavy to handle, and that (physical) fields
are usually considered to be of tensorial nature in  some
representation space of  a symmetry, it is first necessary for the BRS 
algebra presentation of $\cW$-symmetry to switch  from the jet-ghosts
to tensor ones.   
 This will make some contact with the results on the subject disseminated 
through the literature 
\cite{deBoer:1996nu,DiFrancesco:1991qr,Bilal:1991wn,Gov95,Gov96}.  
This kind of problem is often encountered in the treatment of a local
field theory, and even in the B.R.S.T. quantization scheme, see
e.g. \cite{Barnich:2000zw,Brandt:2001tg}. 
The solution to this problem is obviously not unique since a tensor is
defined up to a change of basis among tensors. This is why it must be
solved for the moment with the tools at hand. 

Let us consider the hypothesis where the $\cK$'s are {\em not}
universal, except the top one.
Given a level $s$ of truncation, consider the hierarchy of all lower
orders of truncation $j+1\leq s$ which come into the game with their
own structure functions. At the level $s$, if $\vec{Z}$ denotes a
vector in $\mathbb{C} P^{s-1}$, one has respectively for the DOR and
the variation  
\begin{eqnarray}
        \prt_{(s)} \vec{Z}\zbzs = \sum_{\ell=1}^{s-1}
        \cR_{(s)}^{(\ell)}\zbzs \prt_{(\ell)} \vec{Z}\zbzs, \qquad\quad 
        \delta_{\cWs} \vec{Z}\zbzs = \sum_{\ell=1}^{s-1}
        \cK^{(\ell)}\zbzs \prt_{(\ell)} \vec{Z}\zbzs. 
\end{eqnarray}
Suppose now that, the DOR is rather implemented at the sub-level $s-1$
on the $s-1$ scalar fields, with 
\begin{eqnarray}
        \prt_{(s-1)} \vec{Z}\zbzs = \sum_{\ell=1}^{s-2}
        \cR_{(s-1)}^{(\ell)}(z,\bz|s-1) \prt_{(\ell)} \vec{Z}\zbzs, 
\end{eqnarray}
then the above variation becomes can be projected onto that of the level $s-1$
\begin{eqnarray}
        \delta_{\cWs} \vec{Z}\zbzs &=& \sum_{\ell=1}^{s-2} \Big(
        \cK^{(\ell)}\zbzs + \cK^{(s-1)}\zbzs
        \cR_{(s-1)}^{(\ell)}(z,\bz|s-1) \Big) \prt_{(\ell)}
        \vec{Z}\zbzs \nn\\ 
        &\stackrel{!}{=}& \delta_{\cW_{s-1}} \vec{Z}\zbzs =
        \sum_{\ell=1}^{s-2} \cK^{(\ell)}(z,\bz|s-1) \prt_{(\ell)}
        \vec{Z}\zbzs . 
\end{eqnarray}
Upon requiring that $\delta_{\cWs} \vec{Z}\zbzs = \delta_{\cW_{s-1}}
\vec{Z}\zbzs$, one can identify the top tensorial ghost of the level
$s-1$ in terms of those of the upper level $s$ through the structure
functions of the level $s-1$, by 
\begin{eqnarray}
        \cK^{(s-2)}(z,\bz|s-1) = \cK^{(s-2)}\zbzs + \cK^{(s-1)}\zbzs
        \cR_{(s-1)}^{(s-2)}(z,\bz|s-1). 
\end{eqnarray} 
Repeating the DOR from the level $s$ to an arbitrary sub-level $j$, with
$2\leq j\leq s-1$, the requirement that $\delta_{\cWs} \vec{Z}\zbzs =
\delta_{\cW_j} \vec{Z}\zbzs$ yields for each top tensorial ghost
\begin{eqnarray}
        \cK^{(j-1)}(z,\bz|j) &=& \cK^{(j-1)}\zbzs + \sum_{m=j}^{s-1}
        \cK^{(m)}\zbzs \cR_{(m)}^{(j-1)}(z,\bz|j) \nn\\ 
        &=& \sum_{m=j-1}^{s-1} \cK^{(m)}\zbzs \cR_{(m)}^{(j-1)}(z,\bz|j).
\end{eqnarray}
Next, taking the the ghost of highest conformal weight in each of the
sub-algebras, 
one generates a hierarchy of $j$-contravariant conformal tensors as 
\begin{eqnarray}
        \cC^{(j)}\zbz := \cK^{(j)}(z,\bz|j+1) \ .
\lbl{Cj}
\end{eqnarray}
All the above considerations suggest to take as an ansatz for the
$j$-contravariant ghost conformal tensors the following pretty tricky
linear combination 
\begin{eqnarray}
\cC^{(j)}\zbz = \sum_{m=j}^{s-1}\cK^{(m)}\zbzs \cR_{(m)}^{(j)}(z,\bz|j+1), 
\qquad \mbox{for } 1\leq j\leq s-1,
\lbl{solal}
\end{eqnarray}
where the lower orders of truncation (implemented by the
DOR$_{s-1\rightarrow j}$) crucially enter the construction. Due to
 the requirement that $\delta_{\cWs} \vec{Z}\zbzs =
\delta_{\cW_k} \vec{Z}\zbzs$, for $k=2,\dots,s$, it is worthwhile to
emphasize that the tensorial ghosts $\cC^{(j)}$ carry an universal
nature (regarding all the hierarchy of $\cW_j$-algebras), in the
sense that for $j=1,\dots,k-1$, $k\geq j+1$ one has the identity 
\begin{eqnarray}
\cC^{(j)}\zbz = \sum_{m=j}^{s-1}\cK^{(m)}\zbzs
\cR_{(m)}^{(j)}(z,\bz|j+1) = \sum_{m=j}^{k-1} \cK^{(m)}(z,\bz|k)
\cR_{(m)}^{(j)}(z,\bz|j+1) . 
\lbl{solal'} 
\end{eqnarray}
The latter (which generalizes \rf{solal}) shows that the $\cC^{(j)}$'s do
not depend on the level $k$ of truncation for $k\geq j$. This strongly suggests
that the tensorial ghosts $\cC^{(j)}$'s are of universal nature. Moreover,
one gets for the DOR$_{s-1\rightarrow k-1}$ and with $\ell =
1,\dots,k-1$ 
\begin{eqnarray}
\cK^{(\ell)}(z,\bz|k) = \cK^{(\ell)}\zbzs + \sum_{m=k}^{s-1}
\cK^{(m)}\zbzs \cR_{(m)}^{(\ell)} (z,\bz|k) = \sum_{m=1}^{s-1}
\cK^{(m)}\zbzs \cR_{(m)}^{(\ell)}(z,\bz|k).
\lbl{Ksub}
\end{eqnarray}
Next, by comparing in order to guarantee the transitivity property,
DOR$_{s-1\rightarrow j}$ with DOR$_{s-1\rightarrow k-1}$ followed by
DOR$_{k-1\rightarrow j}$, the structure functions must verify for any
$m \geq k$ with $j+1\leq k \leq s-1$ 
\begin{eqnarray}
        \cR_{(m)}^{(j)}(z,\bz|j+1) = \sum_{\ell = j}^{k-1}
        \cR_{(m)}^{(\ell)} (z,\bz|k) \cR_{(\ell)}^{(j)}(z,\bz|j+1), 
        \lbl{Rlink}
\end{eqnarray}
an identity which has to be checked with the help of \rf{diff_inv} and both
the choices of $j$ and of $k-1$ coordinates among the $s-1$
coordinates given by $\vec{Z}\in\mathbb{C}P^{s-1}$ (identity between
determinants). These two choices of sub-coordinates define
sub-manifolds in $\mathbb{C}P^{s-1}$.  
This phenomenon is the signature of the presence of flag manifolds
denoted by $F_{j\cdots s-2}\mathbb{C}^{s-1}$ over
$\mathbb{C}P^{s-1}$, a geometric concept already mentioned as related
to $\cW$-algebras in \cite{Bilal:1991wn}. In particular the choice
$j=1$ and $k=s$ in 
\rf{Rlink} corresponds to the whole hierarchy \rf{solal} of the $\cC$
ghosts and is associated to the flag manifold $F_{12\cdots
  s-2}\mathbb{C}^{s-1}$. 

Owing to the above considerations, let us define now, for a fixed
level $s$, the following nilpotent operator 
${\dps \delta_{\cWs} = \bigoplus_{\ell = 2}^{s-1} \delta_{\cW_\ell} }$, with
$\delta_{\cW_\ell}^2=0$ and $\lbrace \delta_{\cW_k},\delta_{\cW_\ell}
\rbrace = 0$, which is in some sense filtrated by the various
sub-DOR's relative to the flag sub-manifold $F_{12\cdots
  s-2}\mathbb{C}^{s-1}$. 
Then the task is to figure out the variations $\delta_{\cWs}
\cC^{(j)}$ for $j=1,\dots, s-1$ in terms of the tensorial $\cC$'s themselves.

Note that the tensorial ansatz \rf{solal} gives an universal
character to each of the tensorial top ghosts
$C^{(\ell-1)}\zbz :=\cK^{(\ell-1)}(z,\bz|\ell)$ of each sub-levels.
By virtue of \rf{Ksub}, the latter linearly depend on both the jet ghosts
$\cK(z,\bz|\ell)$'s of the 
top order $s$ of truncation and the structure functions relative to
the various truncations up to order $s-1$. 

In the course of the checking that the $\cC^{(j)}$'s are indeed
$j$-contravariant conformal tensors, \rf{changeK}, \rf{changeR} and the
identities \rf{Ids} were repeatedly used. The tensor character of
$\cC^{(j)}$ is secured by the choice of
$\cR_{(n)}^{(j)}(z,\bz|j+1)$ with maximum upper index $j$ relatively to
the truncation of level $j+1$. 
This is possible if one picks up these objects from the whole
underlying D.O.R. decompositions  with a truncation mechanism at each
level lower than $s$. 
So the price to pay is the introduction of all the
$\cR_{(n)}^{(j)}(z,\bz|j+1)$ coefficients relative to all the
(sub-)truncations from $j=1$ to $j=s-1$. The latter could have been
implemented by a hierarchy of differential equations of the type
\rf{lr}. The ansatz \rf{solal} is a linear system in a Gauss form
which is easily inverted as
\begin{eqnarray}
  \cK^{(\ell)}(z,\bz|s) &=& 
\sum_{m=\ell}^{s-1}
  \cC^{(m)}\zbz\, \cU_{(m)}^{(\ell)}(z,\bz|\ell+1,\dots,s-1),\qquad \ell =
  1,\dots,s-2 \nn\\[-2mm]
&&\lbl{invU}\\[-2mm]
\cK^{(s-1)}(z,\bz|s) &=& \cC^{(s-1)}\zbz, \nn
\end{eqnarray}
where $\cU_{(m)}^{(\ell)}(z,\bz|\ell+1,\dots,s-1)$ is the coefficient
of the inverse upper triangular matrix which depends polynomially on structure
functions pertaining to the sub-levels from $\ell + 1$ to $s-1$. 
More explicitly, for $k,\ell = 1,\dots, s-1$
\begin{eqnarray}
\cU_{(k)}^{(\ell)}(z,\bz|\ell+1,\dots,s-1) = 
\left\{ \begin{array}{l}
1 \qquad \mbox{if } k=\ell \\[2mm]
0 \qquad \mbox{if } k < \ell \\[2mm]
{\dps  \sum_{j=1}^{k-\ell} (-1)^j
\biggl[ \prod_{i=1}^j  \cR^{(\ell_i)}_{(k_i)}(z,\bz|\ell_i+1) \biggr]
_{\tiny \left\arrowvert \begin{array}{l}
{\tiny
\ell_1 = \ell,\, k_j=k} \\
{\tiny
k_i > \ell_i}\\ 
{\tiny
k_i = \ell_{i+1} \mbox{ for } k-\ell\geq 2} 
\end{array}  \right. } }
\ \mbox{if } k > \ell
\end{array}
\right.
\lbl{U}
\end{eqnarray}

The variation of the $\cC^{(j)}$'s given by \rf{solal} is computed upon
using the variation \rf{brsK} for the 
level $s$ and the variation \rf{sR} for all the sub-levels up to $s-1$
owing to the DOR filtrations. It writes
\begin{eqnarray}
  \delta_{\cW_s} \cC^{(j)}\zbz = \sum_{\ell = j}^{s-1} \left( \delta_{\cW_s}
  \cK^{(\ell)}(z,\bz|s) \cR^{(j)}_{(\ell)}(z,\bz|j+1) - \cK^{(\ell)}(z,\bz|s)
  \delta_{\cW_{j+1}} \cR^{(j)}_{(\ell)}(z,\bz|j+1) \right).
\lbl{sol0}
\end{eqnarray}
and after some algebra we get the following variation,
\begin{eqnarray}
  \delta_{\cW_s} \cC^{(j)}\zbz &=&
\sum_{n=1}^{s-1} \sum_{a=n}^{s-1} \sum_{b=1}^{s-1} \sum_{r=0}^n
\cC^{(a)}\zbz \prt_{(r)} \cC^{(b)}\zbz \,
\cU_{(a)}^{(n)}(z,\bz|a+1,\dots,s-1)
\sum_{\ell=1}^b \sum_{k=r}^n \binom{k}{r} \binom{n}{k}
\nn\\
&& \hskip 1cm \times\
\prt_{(k-r)} \cU_{(b)}^{(\ell)}(z,\bz|b+1,\dots,s-1)
\left(\sum_{p=j}^{s-1}
  \cR_{(n+\ell-k)}^{(p)}(z,\bz|s)\cR_{(p)}^{(j)}(z,\bz|j+1) \right)
\nn\\
&& \hskip -1cm
+\ \sum_{n=j}^{s-1} \sum_{a=n}^{s-1} \sum_{b=1}^j 
\bigg[\sum_{u=1}^j \sum_{r=0}^u 
\cC^{(a)}\zbz \prt_{(r)} \cC^{(b)}\zbz\,
\cU_{(a)}^{(n)}(z,\bz|a+1,\dots,s-1)
\sum_{\ell=1}^b \sum_{k=r}^u \binom{k}{r} \binom{u}{k}
\nn\\
&& \hskip 1cm \times\
\prt_{(k-r)} \cU_{(b)}^{(\ell)}(z,\bz|b+1,\dots,j)
\cR_{(n)}^{(u)}(z,\bz|j+1)\cR_{(u+\ell-k)}^{(j)}(z,\bz|j+1)
\nn\\
&& -\ \sum_{r=0}^n 
\cC^{(a)}\zbz \prt_{(r)} \cC^{(b)}\zbz\,
\cU_{(a)}^{(n)}(z,\bz|a+1,\dots,s-1)
\sum_{\ell=1}^b \sum_{k=r}^n \binom{k}{r}
\nn\\
&& \hskip 1cm \times\
\prt_{(k-r)} \cU_{(b)}^{(\ell)}(z,\bz|b+1,\dots,j)
\cR_{(n+\ell-k)}^{(j)}(z,\bz|j+1)
\bigg]\ .
\lbl{Walgebra}
\end{eqnarray}
Equation \rf{Walgebra} can be disassembled into:
\begin{eqnarray}
\delta_{\cW_s} \cC^{(j)}\zbz&\equiv& \sum_{n=1}^{j} n\, \cC^{(n)}\zbz
  \prt\cC^{(j-n+1)}\zbz + \cX^{(j)}(z,\bz|j+1,\dots,s),
\lbl{chi}
\end{eqnarray}
where the first summand looks like the variation coming from a symplectic
approach \cite{Bandelloni:1999et} to $\cW$-algebra.

It is clear that the last term $\cX$ 
is related to the whole symmetry in the sense that it is a
tensorial differential expression of ghost grading two in the various
structure functions $\cR$'s of the sub-levels. 
Due to the nilpotency of the $\delta_{\cWs}$ BRS operation,
the $\cX^{(j)}$'s defined in \rf{chi} do transform according to
\begin{eqnarray}
\delta_{\cWs} \cX^{(j)}(z,\bz|j+1,\dots,s) &=& \sum_{n=1}^j n \biggl(
 \cC^{(n)}\zbz \prt \cX^{(j-n+1)} (z,\bz|j+1,\dots,s) \nn\\[-2mm]
&&\lbl{s_chi} \\[-2mm]
&& -\   \cX^{(n)}(z,\bz|j+1,\dots,s) \prt 
 \cC^{(j-n+1)}\zbz \biggr). \nn
\end{eqnarray} 
The full completion of the last equation \rf{s_chi} amounts to
introducing (together with their $\cW_s$-variations) all a set of
primary fields ($\cW$-currents) which belong to the tower of all the
nested sub-algebras according to the DOR filtration and the respective
variations for each sub-levels. This provides a general solution for
any $j$, and is, according to   our opinion, the most general explicit
expression given, up to now, for {\it any} $\cW$-algebra in a B.R.S
setting.  Of course, the generic expression \rf{Walgebra} contains the
$\cR$ reduction coefficients of {\it all} the  Forsyth sub-frames.  
However, the variations given by Eq.\rf{Walgebra} do not
generally coincide with the   ones found in the literature: this fact
will be illustrated in the next two examples.  In order to recover the
familiar expressions~\cite{Garajeu:1995jn,Watts98}  nontrivial
redefinitions of the tensorial ghosts involving derivative terms must
be performed. The ansatz \rf{invU} relating the jet ghosts to the
tensorial ones is recast into the form~\cite{BaLa04}:
\begin{eqnarray}
\mbox{for } \ell=1,\dots,s-2,\qquad
\cK^{(\ell)} \zbzs = \cC^{(\ell)} \zbz +
\sum_{p=\ell+1}^{s-1}\ \sum_{r=0}^{p-\ell} \prt_{(r)} \cC^{(p)}\zbz
\cT^{(r,\ell)}_{(p)}\zbzs, 
\lbl{cT}
\end{eqnarray}
where derivatives in the tensorial ghosts explicitly enter and where
$\cT^{(b,m)}_{(p)}\zbzs$ depends only on the structure functions
$\cR_{(s)}^{(\ell)}\zbzs$, $\ell = 1,\dots,s-1$ with
$\cT^{(r,m)}_{(p)}\zbzs = 0$ for $p< m$ and $\cT^{(r, m)}_{(p)}\zbzs =
\delta^{(r)}_{(0)}$ for $p=m$.

In this case, the  new $\cX^{(j)}$'s will
depend only on the structure functions of the level $s$.
So, we stress, that this is a consequence of the technical difficulties
coming from the jets to tensor reduction, 
and it is not a problem of the $\cW$-symmetry itself.

In fact this is deduced by searching the algebraic conditions 
(in terms of ghosts and their derivatives, considered as independent fields),
which put to zero, in all the equations \rf{Walgebra} all the terms
containing   
the structure functions of the sub-levels $\cR(z,\bz|j)$, for $j<s$.
For $s$ of reasonable order, we find that the number of vanishing
conditions allows  
to get an unique solution, and numerically provides the $\cW$ examples
found in the literature.

The present upshot greatly
improves some previous work \cite{BaLa01,Bandelloni:2000ri} in the sense
that it is now possible to construct explicitly the $\cX$-term which
breaks by truncation the $\cW_\infty$-symmetry governed by an
underlying symplectomorphism symmetry \cite{Bandelloni:1999et,BaLa02} to a
finite $\cW_s$-algebra. 
 
To distinguish in a clear way a realization of such a $\cW_s$-structure,
the $\cW_3$ and $\cW_4$ cases will be next computed in great details.
Then, despite the lack of well settled examples in the literature for
 (even if some examples exist \cite{Hornfeck:1992tm,Zhu:1993mc})
  $\cW_s$ ($s\geq 5$), remarkable results in \cite{Gov96} will allow
  to find out a general setting.

\subsection{The $\cW_3$ example}

According to the general construction, one has with $s=3$ as top
level, the two tensorial ghosts 
\begin{eqnarray}
\cC^{(1)}\zbz &=& \cK^{(1)}(z,\bz|2) = \cK^{(1)}(z,\bz|3) + \cK^{(2)}(z,\bz|3) \cR^{(1)}_{(2)}(z,\bz|2)\nn\\
\cC^{(2)}\zbz &=& \cK^{(2)}(z,\bz|3)
\end{eqnarray}
and their holomorphically covariant variations according to the DOR
filtration read 
\begin{eqnarray}
\delta_{\cW_3} \cC^{(2)}\zbz
&=& \cC^{(1)}\zbz\prt\cC^{(2)}\zbz + 2\cC^{(2)}\zbz\prt\cC^{(1)}\zbz + \cC^{(2)}\zbz\prt_{(2)}\cC^{(2)}\zbz \nn\\
&&\qquad +\ \cC^{(2)}\zbz\prt \cC^{(2)}\zbz 
\Big[2\cR^{(2)}_{(3)}(z,\bz|3)-3\cR^{(1)}_{(2)}(z,\bz|2)\Big]
\lbl{dc2} \\[2mm]
\delta_{\cW_3} \cC^{(1)}\zbz
&=& \cC^{(1)}\zbz\prt\cC^{(1)}\zbz+2\cC^{(2)}\zbz\prt\cC^{(2)}\zbz \Big[
\cR^{(1)}_{(3)}(z,\bz|3) -\prt\cR^{(1)}_{(2)}(z,\bz|2) \nn\\
&&\qquad +\ \cR^{(2)}_{(3)}(z,\bz|3)\cR^{(1)}_{(2)}(z,\bz|2)
-\big(\cR^{(1)}_{(2)}(z,\bz|2)\big)^2 \Big] ,
\lbl{dc1}
\end{eqnarray}
where in the course of the calculation eq.\rf{prop_1} has been used.
Performing the holomorphically covariant change of generators
\begin{eqnarray}
\cC^{(1)}\zbz \rightarrow \tld\cC^{(1)}\zbz := \cC^{(1)}\zbz +
\sm{1}{2}\prt\cC^{(2)}\zbz 
+ \cC^{(2)}\zbz \Big[\sm{2}{3}\cR^{(2)}_{(3)}(z,\bz|3)
- \cR^{(1)}_{(2)}(z,\bz|2)\Big],  
\lbl{C'}
\end{eqnarray}
allows to removing the $\cR^{(1)}_{(2)}(z,\bz|2)$ dependence and
\rf{invU} rewrites 
\begin{eqnarray}
\cK^{(1)}(z,\bz|3) &=& \tld\cC^{(1)}\zbz - \sm{1}{2}\prt\cC^{(2)}\zbz
- \sm{2}{3}\cC^{(2)}\zbz\cR^{(2)}_{(3)}(z,\bz|3) \nn\\
\cK^{(2)}(z,\bz|3) &=& \cC^{(2)}\zbz,
\lbl{KC_3}
\end{eqnarray}
and depends only on the level $s=3$.
Next, we get the well known transformations \cite{Garajeu:1995jn,BaLa01} upon
redefining $\tld\cC^{(2)} = \sm{1}{2}\, \cC^{(2)}$ by a numerical rescaling
\begin{eqnarray}
\delta_{\cW_3} \tld\cC^{(1)}\zbz &=& \tld\cC^{(1)}\zbz\prt\tld\cC^{(1)}\zbz
- \sm{4}{3}\cT_{(2)}(z,\bz|3)\, \cC^{(2)}\zbz\prt \cC^{(2)}\zbz \nn\\
&&\qquad +\ \sm{1}{4}\, \biggl(
\prt\cC^{(2)}\zbz\prt^2\cC^{(2)}\zbz
- \sm{2}{3}\cC^{(2)}\zbz\prt^3\cC^{(2)}\zbz\biggr)
\lbl{sc1} \\[2mm]
\delta_{\cW_3}
\cC^{(2)}\zbz &=& \tld\cC^{(1)}\zbz\prt\cC^{(2)}\zbz +
2\cC^{(2)}\zbz\prt\tld\cC^{(1)}\zbz 
\nn
\end{eqnarray}
where the expected combination \rf{projconnec} which introduces into the
game a projective connection, is recovered for the case~$s=3$, 
\begin{eqnarray}
\cT_{(2)}(z,\bz|3) = \sm{1}{2} \biggl( \prt\cR^{(2)}_{(3)}(z,\bz|3) 
- \sm{1}{3} \big(\cR^{(2)}_{(3)}(z,\bz|3)\big)^2 - \cR^{(1)}_{(3)}(z,\bz|3)
\biggr)= \mbox{$\sm{1}{2}$} a_{(2)}^{(3)}\zbz ,
\end{eqnarray}
an expression which depends on the level $s=3$ only. The remarkable
fact in the course of the computation of \rf{sc1} is that all the
quantities pertaining to the sub-order $2$ of truncation have
disappeared thanks to the change of generators \rf{C'} in the
tensorial sector. 

Comforted into this approach, one can compute also
the variation of $\cT_{(2)} (z,\bz|3)$. After a straightforward but
rather tedious calculation one successively obtains
\begin{eqnarray}
  \delta_{\cW_3} \cT_{(2)} (z,\bz|3) &=& \bigg(\prt_{(3)}\cK^{(1)} +
   2\, \cT_{(2)} \prt \cK^{(1)} + \cK^{(1)} \cT_{(2)} \bigg)(z,\bz|3) \nn\\ 
&&\hskip -3cm + \ \bigg( \cK^{(2)} \Big[ \prt_{(3)} \cR^{(2)}_{(3)} +
  \prt_{(2)} \cR^{(1)}_{(3)} + \prt_{(2)}
  \big(\cR^{(2)}_{(3)} \big)^2 
- \sm{2}{3}\cR^{(2)}_{(3)} \prt_{(2)}\cR^{(2)}_{(3)}
- \sm{2}{3}\cR^{(2)}_{(3)} \prt \big(\cR^{(2)}_{(3)} \big)^2 
- 2 \cR^{(1)}_{(3)} \prt \cR^{(2)}_{(3)}
- \sm{4}{3} \cR^{(2)}_{(3)} \prt \cR^{(1)}_{(3)} \Big] \nn\\
&& +\ \prt \cK^{(2)} \Big[ 4\prt_{(2)} \cR^{(2)}_{(3)}
+ \prt\cR^{(1)}_{(3)} + \prt \big(\cR^{(2)}_{(3)} \big)^2
- \sm{2}{3}\big(\cR^{(2)}_{(3)} \big)^3
- \sm{7}{3} \cR^{(2)}_{(3)} \cR^{(1)}_{(3)} \Big] \\
&& +\ \prt_{(2)} \cK^{(2)} \Big[ 5 \prt \cR^{(2)}_{(3)} -
   \cR^{(1)}_{(3)} - \sm{1}{3} \big(\cR^{(2)}_{(3)} \big)^2 \Big] 
+ \sm{4}{3} \cR^{(2)}_{(3)} \prt_{(3)} \cK^{(2)} 
+ \prt_{(4)} \cK^{(2)} \bigg)(z,\bz|3), \nn
\end{eqnarray}
in terms of the two $\cK$ ghosts and the structure functions  relative
to the level $s=3$. According  to \rf{KC_3}, this variation can be
re-expressed in terms of the two tensorial $\tld\cC$ ghosts as
\begin{eqnarray}
  \delta_{\cW_3} \cT_{(2)} (z,\bz|3) &=& \prt_{(3)}\tld\cC^{(1)}\zbz + 2\,
   \cT_{(2)}(z,\bz|3)  
   \prt \tld\cC^{(1)}\zbz + \tld\cC^{(1)}\zbz \cT_{(2)}(z,\bz|3)
    \nn\\[2mm]  
&& \hskip -2cm -\ 2\, \tld\cC^{(2)}\zbz \prt \Big[ \sm{1}{6} \prt_{(2)}\cR^{(2)}_{(3)}
- \sm{1}{6} \prt \big(\cR^{(2)}_{(3)} \big)^2
+ \sm{1}{3} \cR^{(2)}_{(3)} \cR^{(1)}_{(3)}
+ \sm{2}{27}\big(\cR^{(2)}_{(3)} \big)^3
- \sm{1}{2} \prt\cR^{(1)}_{(3)} \Big](z,\bz|3) \\[2mm]
&&\hskip -2cm -\ 3\, \prt \tld\cC^{(2)}\zbz \Big[ \sm{1}{6}
   \prt_{(2)}\cR^{(2)}_{(3)} 
- \sm{1}{6} \prt \big(\cR^{(2)}_{(3)} \big)^2
+ \sm{1}{3} \cR^{(2)}_{(3)} \cR^{(1)}_{(3)}
+ \sm{2}{27}\big(\cR^{(2)}_{(3)} \big)^3
- \sm{1}{2} \prt\cR^{(1)}_{(3)} \Big](z,\bz|3). \nn
\end{eqnarray}
The expression between the brackets corresponds to the associated
$W_3$-current as a cubic differential relative to the level $s=3$, (up
to a factor)
\begin{eqnarray}
  8\, W_{(3)}(z,\bz|3) &=& \Big( \sm{1}{6}
   \prt_{(2)}\cR^{(2)}_{(3)} 
- \sm{1}{6} \prt \big(\cR^{(2)}_{(3)} \big)^2
+ \sm{1}{3} \cR^{(2)}_{(3)} \cR^{(1)}_{(3)}
+ \sm{2}{27}\big(\cR^{(2)}_{(3)} \Big)^3
- \sm{1}{2} \prt\cR^{(1)}_{(3)} \big)(z,\bz|3) 
\lbl{W3_3}\\
&=& \Big( \sm{1}{2} \prt a^{(3)}_{(2)} - a^{(3)}_{(3)} \Big)\zbz, \nn
\end{eqnarray}
where the $a^{(3)}$'s were given in \rf{case_3}.

To sump up, the general conformally covariant differential operator \rf{lr} for
$s=3$ can be recast in terms of the two $\cW$-currents
\begin{eqnarray}
  L_3\zbz = \prt_{(3)} + 2 \cT_{(2)}(z,\bz|3) \prt +
  \prt\cT_{(2)}(z,\bz|3) - 8\, W_{(3)}(z,\bz|3)
\end{eqnarray}
where the last type $(3,0)$-term indicates the difference with the
so-called Bol operator of order 3, see e.g.~\cite{Gieres:1993}.

At the level of the BRS differential algebra, dropping out the $\tld{\
}$ for the tensorial ghosts, 
one has an explicit realization of the so-called principal
$\cW_3$-algebra, related to what it is called the pure $\cW_3$-gravity
\cite{Ooguri:1992by}.
The nilpotent BRS algebra for $\cW_3$ writes in terms of a $\cS$-operation
acting on $\cT$ and $W_{(3)}$ which are of spin 2 and spin 3
$\cW$-currents, respectively and of two conformal ghost fields
$\cC^{(1)},\cC^{(2)}$ 
\begin{eqnarray}
\cS \cC^{(1)} &=& \cC^{(1)}\prt \cC^{(1)} +
\prt\cC^{(2)}\prt^2\cC^{(2)} - \sm{2}{3} \cC^{(2)}\prt^3\cC^{(2)} -
\sm{16}{3} \cT\cC^{(2)}\prt\cC^{(2)} \nn \\[2mm] 
\cS\cC^{(2)}&=& \cC^{(1)}\prt\cC^{(2)} + 2  \cC^{(2)}\prt \cC^{(1)} \nn \\[2mm]
\cS \cT &=& \prt^3\cC^{(1)} + 2\cT \prt \cC^{(1)} + \cC^{(1)}\prt\cT -
8(2\cC^{(2)}\prt W_{(3)} + 3\, W_{(3)}\prt\cC^{(2)})
\lbl{sT3} \lbl{W3alg} \\[2mm]
\cS W_{(3)} &=& \sm{1}{24} \Big( \prt^5\cC^{(2)} + 2\cC^{(2)}\prt^3\cT
+10\cT\prt^3\cC^{(2)} 
+15\prt\cT\prt^2\cC^{(2)} + 9\prt^2\cT\prt\cC^{(2)}\nn\\
&&\hs{10}+\; 16\cT\prt\cT\cC^{(2)} + 16\cT^2\prt\cC^{(2)} \Big)
+ \cC^{(1)}\prt W_{(3)} + 3\,W_{(3)}\prt \cC^{(1)} \ .\nn
\end{eqnarray}
However, at the practical level, the $\cW_3$ case
stands as a particular example in the sense that the change of
generators emerges by itself. But for the instance of $\cW_4$, it is not
evident at first sight, to figure out which change of generators for
$C^{(1)}$ and 
$C^{(2)}$ must be performed. For the moment, there is no a general
criterion at our disposal giving any guidance on that
step. Nevertheless, an explicit realization of the  so-called
principal $\cW_4$-algebra can be constructed along both the ideas  of
respecting the covariance and the dependence of the top level $s=4$
only. These two main ideas are the crux of all the construction and
must be explained in a more geometric setup.

\subsection{The $\cW_4$ case}

According to the general construction, this time one has with $s=4$ as
top level the three tensorial ghosts 
\begin{eqnarray}
\cC^{(1)}\zbz &=& \cK^{(1)}(z,\bz|2) = \cK^{(1)}(z,\bz|4) +
\cK^{(2)}(z,\bz|4) \cR^{(1)}_{(2)}(z,\bz|2) + \cK^{(3)}(z,\bz|4)
\cR^{(1)}_{(3)}(z,\bz|2) \nn\\ 
\cC^{(2)}\zbz &=& \cK^{(2)}(z,\bz|3) = \cK^{(2)}(z,\bz|4) + \cK^{(3)}(z,\bz|4)
\cR^{(2)}_{(3)}(z,\bz|3)
\lbl{C_4}\\
\cC^{(3)}\zbz &=& \cK^{(3)}(z,\bz|4). \nn
\end{eqnarray}
The holomorphically covariant variation according to the DOR
filtration of the top ghost is found to be
\begin{eqnarray}
\delta_{\cW_4} \cC^{(3)}\zbz &=&  \big( \cC^{(1)} \prt\cC^{(3)} + 3
\cC^{(3)} \prt \cC^{(1)} + 2 \cC^{(2)}\prt\cC^{(2)}
+ \cC^{(3)} \prt_{(3)}\cC^{(3)} + 3 \cC^{(3)} \prt_{(2)} \cC^{(2)} 
+ \cC^{(2)} \prt_{(2)} \cC^{(3)} \big)\zbz \nn\\
&& \hskip -2cm
 +\ (\cC^{(3)} \prt \cC^{(3)})\zbz \Big[ 
3 \prt \cR_{(4)}^{(3)}(z,\bz|4) + 3 \cR_{(4)}^{(2)}(z,\bz|4) + 3
\big(\cR_{(4)}^{(3)}(z,\bz|4)\big)^2 - 6 \prt \cR_{(3)}^{(2)}(z,\bz|3) \nn\\
&&\qquad -\ 4 \prt \cR_{(2)}^{(1)}(z,\bz|2) - 5 \cR_{(3)}^{(2)}(z,\bz|3)
\cR_{(4)}^{(3)}(z,\bz|4) - 4 \big(\cR_{(2)}^{(1)}(z,\bz|2)\big)^2 \nn\\
&&\qquad 
+\ 4 \cR_{(2)}^{(1)}(z,\bz|2) \cR_{(3)}^{(2)}(z,\bz|3) 
+ 2 \big(\cR_{(3)}^{(2)}(z,\bz|3)\big)^2 \Big] \nn\\
&& +\ (\cC^{(3)} \prt_{(2)} \cC^{(3)})\zbz \Big[
3 \cR_{(4)}^{(3)}(z,\bz|4) - 4 \cR_{(3)}^{(2)}(z,\bz|3) \Big] \nn\\
&& +\ (\cC^{(3)} \cC^{(2)})\zbz \Big[ 2 \prt \cR_{(3)}^{(2)}(z,\bz|3) - 3
\prt \cR_{(2)}^{(1)}(z,\bz|2) \Big] \nn\\
&& +\ (\cC^{(2)} \prt \cC^{(3)})\zbz \Big[ 2 \cR_{(4)}^{(3)}(z,\bz|4)
- 2 \cR_{(3)}^{(2)}(z,\bz|3) - \cR_{(2)}^{(1)}(z,\bz|2) \Big] \nn\\
&& +\ (\cC^{(3)} \prt \cC^{(2)})\zbz \Big[ 3 \cR_{(4)}^{(3)}(z,\bz|4)
- 2 \cR_{(3)}^{(2)}(z,\bz|3) - 3 \cR_{(2)}^{(1)}(z,\bz|2) \Big]
\lbl{del_4_C3}
\end{eqnarray}
The general ansatz for the conformally covariant change of ghosts which does
lead to the cancellation of the structure functions of the sub-levels
($s=2,3$) in the above variation \rf{del_4_C3} is given by
\begin{eqnarray}
  \cC^{(2)}\zbz &=& \tld\cC^{(2)}\zbz + H_{(0)}\zbz \prt \cC^{(3)}\zbz +
  H_{(1)}\zbz \cC^{(3)}\zbz, \nn\\[-2mm]
&&\lbl{change_4} \\[-2mm]
\cC^{(1)}\zbz &=& \tld\cC^{(1)}\zbz + F_{(0)}\zbz \prt\tld\cC^{(2)}\zbz +
  F_{(1)}\zbz \tld\cC^{(2)}\zbz + L_{(0)}\zbz \prt_{(2)} \cC^{(3)}\zbz\nn\\
  &&\qquad +\ L_{(1)}\zbz \prt \cC^{(3)}\zbz + L_{(2)}\zbz \cC^{(3)}\zbz. \nn
\end{eqnarray}
Cancellation of the sub-levels in \rf{del_4_C3} gives
\begin{eqnarray*}
H_{(0)}\zbz = - \sm{1}{2},\quad   F_{(0)}\zbz = -1,&&
H_{(1)}\zbz = \cR_{(3)}^{(2)}(z,\bz|3) - \sm{3}{4}
\cR_{(4)}^{(3)}(z,\bz|4) \\
F_{(1)}\zbz = \cR_{(2)}^{(1)}(z,\bz|2) - \sm{1}{2}
\cR_{(4)}^{(3)}(z,\bz|4),&&
L_{(1)}\zbz = \sm{1}{4} \cR_{(4)}^{(3)}(z,\bz|4) - \sm{1}{2}
\cR_{(2)}^{(1)}(z,\bz|2) - \prt L_{(0)}\zbz,
\end{eqnarray*}
and the glueing rules for $\cC^{(1)}$ infers $L_{(0)}\zbz =
\sm{1}{5}$. Plugging these results into \rf{change_4} and inverting \rf{C_4}
the three $\cK$ ghosts of the level $s=4$ are re-expressed as
\begin{eqnarray}
\cK^{(3)}(z,\bz|4) &=& \cC^{(3)}\zbz \nn\\ 
\cK^{(2)}(z,\bz|4) &=& \tld\cC^{(2)}\zbz  -\sm{1}{2}\prt
\cC^{(3)}\zbz
- \sm{3}{4} \cR_{(4)}^{(3)}(z,\bz|4) \cC^{(3)}\zbz
\lbl{KC_4} \\ 
\cK^{(1)}(z,\bz|4) &=& \tld\cC^{(1)}\zbz - \prt\tld\cC^{(2)}\zbz
- \sm{1}{2} \cR_{(4)}^{(3)}(z,\bz|4) \tld\cC^{(2)}\zbz 
+ \sm{1}{5} \prt_{(2)} \cC^{(3)}\zbz\nn\\
&& \hskip -2cm +\, \sm{1}{4} \cR_{(4)}^{(3)}(z,\bz|4) \prt
\cC^{(3)}\zbz + \cC^{(3)}\zbz \Big( L_{(2)}\zbz +
\sm{3}{4}\cR_{(2)}^{(1)}(z,\bz|2)\cR_{(4)}^{(3)}(z,\bz|4) -
\cR_{(3)}^{(1)}(z,\bz|2) \Big). \nn
\end{eqnarray}
Since there are various possibilities to cancel the
sub-levels, the remaining function coefficient $L_{(2)}\zbz$ must be
determined with the help of $\delta_{\cW_4} \tld\cC^{(2)}$ computed
from the second equation of \rf{KC_4} and the known variations at the
level $s=4$ for $\cK^{(3)}(z,\bz|4)$, $\cK^{(2)}(z,\bz|4)$ and
$\cR_{(4)}^{(3)}(z,\bz|4)$ according to the general theory. This step
will secure the nilpotency $\delta_{\cW_4}^2=0$. After lengthy
computations performed with the help of {\tt Mathematica}, one ends
with
\begin{eqnarray*}
L_{(2)}\zbz &=& \sm{12}{25} \prt \cR_{(4)}^{(3)}(z,\bz|4) -
  \sm{3}{25} \big(\cR_{(4)}^{(3)}(z,\bz|4)\big)^2 - \sm{41}{50}
  \cR_{(4)}^{(2)}(z,\bz|4)\\
&& -\,
  \sm{3}{4}\cR_{(2)}^{(1)}(z,\bz|2)\cR_{(4)}^{(3)}(z,\bz|4) + 
\cR_{(3)}^{(1)}(z,\bz|2)\ ,
\end{eqnarray*}
which once substituted yields
\begin{eqnarray}
	\cK^{(1)}(z,\bz|4) &=& \tld\cC^{(1)}\zbz - \prt\tld\cC^{(2)}\zbz
- \sm{1}{2} \cR_{(4)}^{(3)}(z,\bz|4) \tld\cC^{(2)}\zbz 
+ \sm{1}{5} \prt_{(2)} \cC^{(3)}\zbz 
\lbl{KC_4_bis} \\
&& \hskip -2cm +\, \sm{1}{4} \cR_{(4)}^{(3)}(z,\bz|4) \prt
\cC^{(3)}\zbz + \cC^{(3)}\zbz \Big( \sm{12}{25} \big( \prt \cR_{(4)}^{(3)}(z,\bz|4) -
  \sm{1}{4} \big(\cR_{(4)}^{(3)}(z,\bz|4)\big)^2 \big)- \sm{41}{50}
  \cR_{(4)}^{(2)}(z,\bz|4) \Big).\nn 
\end{eqnarray}
This shows that the system \rf{invU} can be rewritten
in terms of the structure functions of the level $s=4$ only thanks to
the redefinition \rf{change_4} of the tensorial ghosts coming from the
sub-levels. 
Recall that these redefinitions are required for re-absorbing 
the structure functions of the sub-levels. The change of generators \rf{KC_4}
also confirms the general ansatz \rf{cT} given in \cite{BaLa04}. 
The variation \rf{del_4_C3} then reduces to
\begin{eqnarray}
\hskip -1cm
\delta_{\cW_4} \cC^{(3)}\zbz &=& \tld\cC^{(1)}\zbz \prt\cC^{(3)}\zbz + 3\,
\cC^{(3)}\zbz \prt \tld\cC^{(1)}\zbz + 2\,
\tld\cC^{(2)}\zbz\prt\tld\cC^{(2)}\zbz \nn\\ 
&& \hskip -2cm
+\, \sm{1}{10} \left( \cC^{(3)}\zbz \prt_{(3)}\cC^{(3)}\zbz - 2\, \prt \cC^{(3)}\zbz \prt_{(2)}\cC^{(3)}\zbz
+ 14\, \cT_{(2)}(z,\bz|4)\cC^{(3)}\zbz \prt \cC^{(3)}\zbz \right)
\lbl{sC3}
\end{eqnarray}
from which emerges the projective connection $\cT_{(2)}(z,\bz|4)$ associated to the level $s=4$,
\begin{eqnarray}
\cT_{(2)}(z,\bz|4) := \Big( \sm{3}{10} \prt \cR_{(4)}^{(3)}
- \sm{3}{40} \big(\cR_{(4)}^{(3)} \big)^2
  - \sm{1}{5} \cR_{(4)}^{(2)} \Big)(z,\bz|4) = \sm{1}{5}\,
  a_{(2)}^{(4)}\zbz
\lbl{T4}
\end{eqnarray}
where $a_{(2)}^{(4)}$ given by \rf{projconnec} carries the projective
connection property as it was already checked by using the
general glueing rules \rf{bdle}. 
The variation $\delta_{\cW_4} \tld\cC^{(2)}$ is then computed to be
\begin{eqnarray}
\delta_{\cW_4} \tld\cC^{(2)}\zbz &=&  \tld\cC^{(1)}\zbz
\prt\tld\cC^{(2)}\zbz + 2\, \tld\cC^{(2)}\zbz \prt \tld\cC^{(1)}\zbz -
\sm{3}{32}\, \cC^{(3)}\zbz \prt\cC^{(3)}\zbz W_{(3)}(z,\bz|4) \nn\\
&&\hskip - 2cm
-\, \sm{1}{10} \left( \tld\cC^{(2)}\zbz\prt_{(3)}\cC^{(3)}\zbz - 3\,
  \prt\tld\cC^{(2)}\zbz \prt_{(2)}\cC^{(3)}\zbz + 5\,
  \prt_{(2)}\tld\cC^{(2)}\zbz \prt\cC^{(3)}\zbz \right.\nn\\
&& \hskip - 2cm
-\ 5\, \prt_{(3)}\tld\cC^{(2)}\zbz \cC^{(3)}\zbz + \big(18\,
  \tld\cC^{(2)}\zbz \prt\cC^{(3)}\zbz - 34\, \prt \tld\cC^{(2)}\zbz
  \cC^{(3)}\zbz \big)\cT_{(2)}(z,\bz|4)\nn\\
&&\hskip - 2cm\left.  -\ 7\, \tld\cC^{(2)}\zbz
  \cC^{(3)}\zbz \prt\cT_{(2)}(z,\bz|4) \right)
\lbl{sC2}
\end{eqnarray}
where one also gets the conformally covariant spin three $\cW$-current
associated to the top level $s=4$, 
\begin{eqnarray}
W_{(3)}(z,\bz|4) &:=& \Big( 8 \prt \cR_{(4)}^{(2)} - 4 \prt_{(2)}
\cR_{(4)}^{(3)} + 3 \prt \big(\cR_{(4)}^{(3)} \big)^2
- 4 \cR_{(4)}^{(2)} \cR_{(4)}^{(3)} - 8
\cR_{(4)}^{(1)} \Big) (z,\bz|4) .
\lbl{W3_4}
\end{eqnarray}
Going on through the computation of the variation with the help
of the third equation in \rf{KC_4} and the known variations at the
level $s=4$ for $\cC^{(3)}$, $\tld\cC^{(2)}$, $\cR_{(4)}^{(3)}(z,\bz|4)$ and
$\cR_{(4)}^{(2)}(z,\bz|4)$, and after redefining $\cC^{(3)}:= -
320\,\tld\cC^{(3)}$ by a numerical factor for later convenience, one gets,
\begin{eqnarray}
 \delta_{\cW_4} \tld\cC^{(1)}\zbz &=& (\tld\cC^{(1)}
\prt\tld\cC^{(1)})\zbz + \sm{3}{5} \left( \prt\tld\cC^{(2)}
  \prt_{(2)}\tld\cC^{(2)} - \sm{2}{3}
  \tld\cC^{(2)}\prt_{(3)}\tld\cC^{(2)} - \sm{16}{3}
  \tld\cC^{(2)}\prt\tld\cC^{(2)}\cT_{(2)}(z,\bz|4) \right)\zbz\nn\\
&&+\Big( 20\, \tld\cC^{(2)} \prt\tld\cC^{(3)} - \sm{108}{5}\prt \tld\cC^{(2)}
\tld\cC^{(3)} \Big)\zbz W_{(3)}(z,\bz|4) + \sm{28}{5}
\tld\cC^{(2)}\zbz \tld\cC^{(3)}\zbz \prt W_{(3)}(z,\bz|4) \nn\\
&&+\, 1024 \left[ 3\, \tld\cC^{(3)} \prt_{(5)}\tld\cC^{(3)} - 5\,
  \prt\tld\cC^{(3)} \prt_{(4)}\tld\cC^{(3)} + 6\,
  \prt_{(2)}\tld\cC^{(3)} \prt_{(3)}\tld\cC^{(3)} + 57\, \tld\cC^{(3)}
  \prt_{(2)}\tld\cC^{(3)} \prt\cT_{(2)}(z,\bz|4) \right.\nn\\
&&+ \Big( 78\, \tld\cC^{(3)}
  \prt_{(3)}\tld\cC^{(3)} - 118 \prt\tld\cC^{(3)}
  \prt_{(2)}\tld\cC^{(3)} \Big) \cT_{(2)}(z,\bz|4)\nn\\
&&\left. +\, \tld\cC^{(3)}
  \prt\tld\cC^{(3)} \Big( 57 \prt_{(2)}\cT_{(2)}(z,\bz|4) + 432
  \big(\cT_{(2)}(z,\bz|4)\big)^2 - 14 W_{(4)}(z,\bz|4) \Big)
\right]\zbz,
\lbl{sC1}
\end{eqnarray}
from where emerges a $\cW$-current of spin 4 (a $(4,0)$-type conformally
covariant differential) associated to the level $s=4$,
as it can be checked by using \rf{bdle},
\begin{eqnarray}
  800\, W_{(4)}(z,\bz|4) &=& \Big( 144\,\big( \cR_{(4)}^{(2)} \big)^2 +
  400\,\cR_{(4)}^{(1)} \cR_{(4)}^{(3)} + 208\, \cR_{(4)}^{(2)}\big(
  \cR_{(4)}^{(3)} \big)^2 + 39 \big( \cR_{(4)}^{(3)} \big)^4 - 800\,
  \prt\cR_{(4)}^{(1)}\nn\\
&&  -\, 400\,\cR_{(4)}^{(3)} \prt\cR_{(4)}^{(2)} -
  432\, \cR_{(4)}^{(2)} \prt\cR_{(4)}^{(3)} - 104\, \prt\big(
  \cR_{(4)}^{(3)} \big)^3 + 264\, \big(\prt\cR_{(4)}^{(3)}\big)^2\nn\\
&& +\, 320\, \prt_{(2)}\cR_{(4)}^{(2)} + 240\,
  \cR_{(4)}^{(3)}\prt_{(2)}\cR_{(4)}^{(3)} - 80\,
  \prt_{(3)}\cR_{(4)}^{(3)} \Big) (z,\bz|4).
\lbl{W4_4}
\end{eqnarray}
Hence, the general conformally covariant differential operator \rf{lr} for
$s=4$ expressed in terms of the three $\cW_4$-currents is
\begin{eqnarray}
  L_4\zbz &=& \prt_{(4)} + 5 \cT_{(2)}(z,\bz|4) \prt_{(2)} +
  5\prt\cT_{(2)}(z,\bz|4)\prt + \sm{3}{2}\left(\prt_{(2)}\cT_{(2)}(z,\bz|4)
  + \sm{3}{2} \big( \cT_{(2)}(z,\bz|4) \big)^2 \right)\nn\\
&& +\ \sm{1}{8}\,
  W_{(3)}(z,\bz|4)\prt + \sm{1}{16} \prt W_{(3)}(z,\bz|4) - \sm{1}{2}
  W_{(4)}(z,\bz|4), 
\end{eqnarray}
where the first line is the Bol operator of order 4, see
e.g. \cite{Gieres:1993}, depending only on the projective
connection~$\cT_{(2)}$.  

All this BRS algebra is an explicit realization of the so-called principal
$\cW_4$-algebra for pure $\cW_4$-gravity~\cite{Watts98,Hornfeck:1993hh}.
Performing the rescaling 
\begin{eqnarray*}
  \tld \cC^{(2)} \lra -8i \sqrt{5}\,\tld\cC^{(2)},\qquad W_{(3)}(z,\bz|4)
  \lra \sm{i}{8\sqrt{5}}\,W_{(3)}(z,\bz|4),
\end{eqnarray*}
and dropping out both the $\tld{\ }$ for the tensorial ghosts and the
explicit reference to the the level $s=4$, one gets the
presentation as a full nilpotent BRS algebra for $\cW_4$-algebra, 

\begin{eqnarray}
\cS\, \cC^{(1)} &=& \cC^{(1)}\prt \cC^{(1)} 
- 192 \Big(\prt\cC^{(2)}\prt_{(2)}\cC^{(2)} - \sm{2}{3} 
\cC^{(2)}\prt_{(3)}\cC^{(2)} - \sm{16}{3} \cT\cC^{(2)}\prt\cC^{(2)} \Big) \nn\\
&& +\; 256 \Big(27 \prt \cC^{(2)} \cC^{(3)} W_{(3)} - 25 \cC^{(2)}
\prt\cC^{(3)} W_{(3)}  - 7 \cC^{(2)} \cC^{(3)} \prt W_{(3)} \Big) \nn\\
&& +\; 1024 \Big(3 \cC^{(3)} \prt_{(5)} \cC^{(3)} - 5 \prt \cC^{(3)}
\prt_{(4)} \cC^{(3)} 
+ 6 \prt_{(2)} \cC^{(3)} \prt_{(3)} \cC^{(3)} + 57 \prt_{(2)} \cT
\cC^{(3)} \prt \cC^{(3)} 
\nn\\
&& \hs{10} +\; 57 \prt\cT \cC^{(3)} \prt_{(2)} \cC^{(3)} +
(78 \cC^{(3)} \prt_{(3)}\cC^{(3)} - 118 \prt \cC^{(3)} \prt_{(2)}
\cC^{(3)})\cT\nn\\ 
&&\hs{10} -\; 14 \cC^{(3)} \prt \cC^{(3)} W_{(4)} 
+ 432 \cC^{(3)} \prt \cC^{(3)} \cT^2\Big) \nn\\[2mm] 
\cS\, \cC^{(2)} &=& \cC^{(1)} \prt \cC^{(2)} + 2 \cC^{(2)} \prt \cC^{(1)} 
+ 32 \Big( \cC^{(2)} \prt_{(3)} \cC^{(3)} - 3 \prt \cC^{(2)}
\prt_{(2)} \cC^{(3)} + 5 \prt_{(2)} \cC^{(2)} \prt \cC^{(3)} 
- 5 \prt_{(3)} \cC^{(2)} \cC^{(3)}\nn\\
&& \hs{10} +\; 18 \cC^{(2)} \prt \cC^{(3)} \cT
- 34 \prt \cC^{(2)} \cC^{(3)}  \cT
- 7 \cC^{(2)} \cC^{(3)} \prt \cT \Big)
 - 9600\, \cC^{(3)} \prt \cC^{(3)} W_{(3)} \nn\\[2mm]
\cS\, \cC^{(3)} &=& \cC^{(1)} \prt\cC^{(3)} + 3 \cC^{(3)} \prt \cC^{(1)}
+ 2 \cC^{(2)}\prt\cC^{(2)} \nn\\
&& -\; 32 \Big( \cC^{(3)} \prt_{(3)}\cC^{(3)} - 2 \prt \cC^{(3)}
\prt_{(2)} \cC^{(3)} + 14 \cT \cC^{(3)} \prt \cC^{(3)} \Big)
\nn \\[2mm]
\cS\, \cT &=& \prt_{(3)}\cC^{(1)} + 2\cT \prt\cC^{(1)} + \cC^{(1)} \prt\cT 
- 8 \Big( 2 \cC^{(2)}\prt W_{(3)} + 3 W_{(3)} \prt \cC^{(2)} \Big) \nn\\
&& +\; 32 \Big( 3 \cC^{(3)} \prt W_{(4)} + 4 W_{(4)} \prt \cC^{(3)} \Big)
\lbl{W4alg} \\[2mm]
\cS\, W_{(3)} &=& \cC^{(1)}\prt W_{(3)} + 3\,W_{(3)}\prt \cC^{(1)}
- 8 \Big(\prt_{(5)}\cC^{(2)} + 2\cC^{(2)}\prt_{(3)}\cT
+ 10\cT \prt_{(3)}\cC^{(2)} + 15\prt\cT \prt_{(2)}\cC^{(2)} \nn\\
&& \hs{10} +\;  9\prt_{(2)}\cT\prt\cC^{(2)} +
16\cT\prt\cT \cC^{(2)}
+ 16 \cT^2\prt\cC^{(2)} + \cC^{(2)} \prt W_{(4)} + 2 W_{(4)}
\prt\cC^{(2)} \Big)\nn\\ 
&& +\; 32 \Big( 5 \cC^{(3)} \prt_{(3)} W_{(3)} + 10 \prt \cC^{(3)} \prt_{(2)} W_{(3)} 
+ 28 \prt_{(2)} \cC^{(3)} \prt W_{(3)} + 14 \prt_{(3)} \cC^{(3)} W_{(3)} \nn\\
&&\hs{10} +\; 34 \cC^{(3)} \cT \prt W_{(3)} + 27 \cC^{(3)} W_{(3)} \prt\cT 
+ 52 \prt \cC^{(3)}\cT W_{(3)} \Big) \nn
\\[2mm]
\cS\, \cW_{(4)} &=& \cC^{(1)}\prt W_{(4)} + 4 W_{(4)} \prt \cC^{(1)}
- 8 \Big( \cC^{(2)} \prt_{(3)} W_{(3)} + 6 \prt \cC^{(2)} \prt_{(2)} W_{(3)} 
+ 14 \prt_{(2)} \cC^{(2)} \prt W_{(3)} + 14 \prt_{(3)} \cC^{(2)} W_{(3)}\nn\\ 
&&\hs{10} +\; 18 \cC^{(2)} \cT \prt W_{(3)} + 25 \cC^{(2)} \prt\cT W_{(3)} +
52 \prt\cC^{(2)} \cT W_{(3)} \Big) \nn\\ 
&& +\; 32 \Big( \prt_{(7)} \cC^{(3)} + 3 \cC^{(3)} \prt_{(5)} \cT + 20
\prt \cC^{(3)} \prt_{(4)} \cT + 56 \prt_{(2)} 
\cC^{(3)} \prt_{(3)} \cT + 84 \prt_{(3)} \cC^{(3)} \prt_{(2)} \cT + 70
\prt_{(4)} \cC^{(3)} \prt \cT \nn\\  
&&\hs{10} +\; 28 \prt_{(5)}\cC^{(3)} \cT + \cC^{(3)}( 177\prt\cT\prt_{(2)}\cT 
+ 78 \cT\prt_{(3)}\cT) + \prt\cC^{(3)}(352 \cT\prt_{(2)}\cT + 295 (\prt
\cT)^2)\nn\\   
&&\hs{10} +\; 588 \prt_{(2)}\cC^{(3)} \cT\prt\cT + 196 \prt_{(3)}\cC^{(3)}
\cT^2 + 432 \cC^{(3)} \cT^2\prt\cT + 288 \prt\cC^{(3)}\cT^3 \nn\\
&&\hs{10} +\; 75 \cC^{(3)} W_{(3)}\prt W_{(3)} + 75 \prt\cC^{(3)} (W_{(3)})^2 
- \cC^{(3)} \prt_{(3)} W_{(4)} - 5 \prt\cC^{(3)} \prt_{(2)} W_{(4)} 
- 9 \prt_{(2)}\cC^{(3)} \prt W_{(4)} \nn\\ 
&&\hs{10} -\; 6 \prt_{(3)}\cC^{(3)} W_{(4)} - 14 \cC^{(3)} \prt(\cT W_{(4)}) 
- 28 \prt\cC^{(3)} \cT W_{(4)} \Big)\ .\nn 
\end{eqnarray}
Remind once more that there is a breaking term in the top ghost
variation $\cS C^{(3)}$ with respect to the symplectic variation,
 so that the mechanisms using the so-called $\theta$-trick
described in previous papers
\cite{Bandelloni:1999et,Bandelloni:2000ri} for the $\cW_3$ case
does not work in the $\cW_4$ case. Let us remark that if one sets
$\cC^{(3)}=0$ and $W_{(4)}=0$ 
and performs the rescalings of the generators
$\cC^{(2)}\lra \frac{i\sqrt{3}}{24}\,\cC^{(2)}$ and
$W_{(3)} \lra - 8i\sqrt{3}\,W_{(3)}$ in \rf{W4alg} then the
$\cW_3$-algebra \rf{W3alg} is recovered. This confirms the 
universal definition \rf{Cj} of the tensorial ghosts
as $\cC^{(s-1)}\zbz = \cK^{(s-1)}(z,\bz|s)$ as the top ghost of each level
$s$ and also the interweaving of the algebras dictated by the successive DOR's.

\subsection{Comparison with some previous work} \label{compar} 

The general ansatz \rf{cT} given in \cite{BaLa04} and exemplified in
\rf{KC_3} and \rf{KC_4}, \rf{KC_4_bis} for $s=3,4$ respectively, can
be put into relation with some previous pioneer
work~\cite{DiFrancesco:1991qr,Bilal:1991wn,Gov95,Gov96}. Indeed,
\cite{Gov96} will be of particular interest. There ``Beltrami
differentials'' emerging from a 
multi-time approach for KdV flows were related to ``Bilal-Fock-Kogan''
generalized tensorial Beltrami coefficients \cite{Bilal:1991wn}
appearing in $\cW$-gravity
along the ideas of \cite{DiFrancesco:1991qr}. According to their
contravariant behavior these various type of Beltrami deformation
parameters can be used in order to recover our ansatz \rf{cT}.

As said in the Introduction, working with either homogeneous or
inhomogeneous coordinates seems to be a matter of taste. In our
construction, the latter were preferred because they strengthen the
role of the symmetry algebra.

If one considers the homogeneous solutions $f$ of the $s$-th order
conformally covariant linear equation \rf{lr}, these solutions as
$\left(\sm{1-s}{2}\right)$-conformal fields are 
equivalently subject to a DOR since the $s$-th order derivative can be
expressed in terms of the lower order ones and the smooth coefficient
of the operator $L_s$. Their variation under large chiral 
diffeomorphisms were computed in \cite{Bandelloni:2002zg} to be
\begin{eqnarray} 
\delta_{\cW_s}f\zbz = \sum_{\ell=0}^{s-1}\cM^{(\ell)}\zbzs
\prt_{(\ell)}f\zbz \ .
\lbl{brsf}
\end{eqnarray}
This variation for homogeneous coordinates must be related to the
variation \rf{brsZ} for the inhomogeneous coordinates. 
Indeed Eqs.\rf{ZR}, \rf{recurf1} allow to find a complete link between
the ghosts $\cK^{(m)}\zbzs$ and $\cM^{(\ell)}\zbzs$,
\begin{eqnarray}
\cK^{(m)}\zbzs = \sum_{\ell=m}^{s-1} \binom{\ell}{m} \cM^{(\ell)}\zbzs
\cQ_{(\ell-m)}\zbzs , \qquad m = 1,\dots,s-1
\lbl{KM}
\end{eqnarray}
and gives a direct answer to a problem raised in 
\cite{DiFrancesco:1991qr}  about the $\cW$ deformations  
 of the $f$ functions via the KdV "multi-time" approach,
 providing a direct expression of the KdV hierarchy. 

Inspired by \cite{Gov96}, one can mimic the construction used for
relating KdV flows and $\cW$-diffeomorphisms according to
the following dictionary 
\begin{eqnarray}
  \delta \longleftrightarrow \bprt,\qquad \cM^{(\ell)} \longleftrightarrow
\mu_\ell,\quad \mbox{and} \quad \tld{\cC}^{(k)} \longleftrightarrow
\rho_k ,
\lbl{dictionary}
\end{eqnarray}
where the $\cM$ play the role of the ghost parameters for KdV flows
and $\tld{\cC}$\footnote{For the sake of consistency with the treated
  examples one uses the $\tld{\cC}$ ghosts.} those for the
infinitesimal $\cW$-diffeomorphisms. 

One can conformally covariantize the variation \rf{brsf} by
introducing tensorial ghosts $\tld{\cC}$ which serve to filtrate the
variation by their conformal weight according to
\begin{eqnarray}
\delta_{\cW_s}f = \sum_{k=1}^{s-1}
\mathscr{B}_{(k)}(\tld{\cC}^{(k)},a^{(s)}_{(2)}) f
\lbl{brsfB}
\end{eqnarray}
where the $\mathscr{B}_{(k)}$ are the conformally covariant differential
operators constructed in \cite{Gov96} mapping
$\left(\sm{1-s}{2}\right)$-conformal 
fields into themselves. The coefficient function $a^{(s)}_{(2)}$ has a
prominent role since it is related to a projective connection (see
\rf{projconnec}) and controls the M\"obius transformations.
For $a^{(s)}_{(2)} \equiv 0$
\footnote{Owing to \rf{projconnec} this implies a non trivial
  differential constraint on the  
 structure functions and then a kind of group contraction.} one
recalls that \cite{Gov96} 
\begin{eqnarray}
  \mathscr{B}_{(k)}(\tld{\cC}^{(k)},a^{(s)}_{(2)}\equiv 0) =
  \sum_{j=k}^{s-1} \gamma_{(k)}^{(k-j)}[s] \big( \prt_{(mkj)}
  \tld{\cC}^{(k)} \big) \prt_{(j)} 
\end{eqnarray}
with $\gamma_{(k)}^{(0)}[s] = 1$ fixing the normalization between
$\cM$ and $\tld{\cC}$. Comparison between the variations \rf{brsf} and
\rf{brsfB} yields
\begin{eqnarray}
\cM^{(0)}\zbzs &=& \sum_{k=1}^{s-1} \gamma_{(k)}^{(k)}[s]
   \prt_{(k)} \tld{\cC}^{(k)}\zbz \nn\\ 
\cM^{(\ell)}\zbzs &=& \sum_{k=\ell}^{s-1} \gamma_{(k)}^{(k-\ell)}[s]
\prt_{(k-\ell)} \tld{\cC}^{(k)}\zbz , \qquad \ell = 1,\dots,s-1 
	\lbl{MC}
\end{eqnarray}
where the numerical coefficients were given in \cite{Gov96}
\begin{eqnarray}
\gamma_{(k)}^{(j)}[s] = (-1)^j\,
\frac{\binom{s+j-k-1}{j}\,\binom{k}{j}}{\binom{2k}{j}}, \qquad
\mbox{with } \gamma_{(k)}^{(0)}[s] = 1 
\end{eqnarray}
as solutions of the recursive equation
\begin{eqnarray*}
(j+1)(2k - j)  \gamma_{(k)}^{(j+1)}[s] + (k-j)(s+j-k)
\gamma_{(k)}^{(j)}[s] = 0 \,
\end{eqnarray*}
coming from the study of the covariance of \rf{brsfB} under projective
holomorphic transformations.

Inserting \rf{MC} into \rf{KM} one gets at $a^{(s)}_{(2)} \equiv 0$
\begin{eqnarray}
\cK^{(m)}\zbzs = \sum_{\ell=m}^{s-1} \binom{\ell}{m} 
\cQ_{(\ell-m)}\zbzs \sum_{k=\ell}^{s-1} \gamma_{(k)}^{(k-\ell)}[s]
\prt_{(k-\ell)} \tld{\cC}^{(k)}\zbz, \qquad m = 1,\dots,s-1 
\lbl{KC}
\end{eqnarray}
The dependence in $a^{(s)}_{(2)}$ can be restored by studying the
conformal covariance of \rf{brsfB} under an arbitrary holomorphic
transformation. The change \rf{KC} corresponds to
\begin{eqnarray}
\delta_{\cW_s} Z\zbzs = \sum_{k=1}^{s-1}
\mathscr{D}_{(k)}(\tld{\cC}^{(k)},a^{(s)}_{(2)}) Z\zbzs
\lbl{brsZB}
\end{eqnarray}
an equivalent to \rf{brsfB}. Thanks to filtration by the tensorial ghosts
$\cC^{(k)}$, for each $k$, the operator
$\mathscr{D}_{(k)}(\tld{\cC}^{(k)},a^{(s)}_{(2)})$ acting on scalar fields 
has no constant term (by \rf{brsZ}) and must be a scalar under holomorphic
transformations. They can be obtained by using the $\mathscr{B}$ operators 
computed in \cite{Gov96} without taking into account their constant
terms since the inhomogeneous coordinates $Z$ are
used in the present paper. 
For the sake of completeness, one rewrites the first few of them
\begin{eqnarray}
\mathscr{D}_{(1)}(\tld{\cC}^{(1)},a^{(s)}_{(2)}) &=& \tld{\cC}^{(1)} \prt \nn\\
\mathscr{D}_{(2)}(\tld{\cC}^{(2)},a^{(s)}_{(2)}) &=& \tld{\cC}^{(2)} \prt_{(2)} 
- \sm{s-2}{2} \prt \tld{\cC}^{(2)} \prt \\
\mathscr{D}_{(3)}(\tld{\cC}^{(3)},a^{(s)}_{(2)}) &=& \tld{\cC}^{(3)} \prt_{(3)}
- \sm{s-3}{2} \prt \tld{\cC}^{(2)} \prt_{(2)} + \left( \sm{(s-2)(s-3)}{10}
  \prt_{(2)}\tld{\cC}^{(3)} + \sm{6(3s^2 -7)}{5(s^3-s)} \tld{\cC}^{(3)}
  a^{(s)}_{(2)} \right) \prt \nn
\end{eqnarray}
A direct confrontation of \rf{KC} (in which the
$a^{(s)}_{(2)}$-dependence has been made explicit) with \rf{KC_3} and
\rf{KC_4}, \rf{KC_4_bis} for respectively $s=3,4$ gives a perfect
accord upon using the recursion \rf{recurf1} and the definition
\rf{projconnec}.

The general ansatz \rf{cT} can be thus recovered with the help of
existing results in the literature. But the linear decomposition
\rf{invU} depending on the 
structure functions of all the possible sub-levels shows the origin of
the tensorial ghosts as the highest conformal weighted parameter in
each of the nested sub-algebras governed by the DOR filtration.
 According to the treated examples $\cW_3$
and $\cW_4$-algebras, the appropriate ghost parameters for the
linear $\cW$-diffeomorphisms are those constructed by redefining $\cC^{(\ell)}
\lra \tld{\cC}^{(\ell)}$,  for the intermediate DOR decompositions in
order to re-absorb all 
the structure functions of the sub-levels. It is worthwhile to notice
that the algorithm is performed in a conformally
covariant manner and in the respect of the nilpotency of the
$\cW$-algebra. 

\section{Conclusion and perspective} \label{concl}

Throughout the paper, we have considered conformal differential
operators defined on a Riemann surface whose solutions are homogeneous
coordinates of some complex projective space.  The latter lead to the notion of
Forsyth frames as projective coordinates.
In this context, our main results are:

(i) linear differential order reductions (DOR), see Theorem
\ref{Stamnt1}, determine the structure functions of the large chiral
symmetry algebra. These structure functions are the central objects of
all our construction;

(ii) conformal differential operators can be explicitly constructed
from the given structure functions entering the linear DOR;

(iii) the
extension to the chiral truncated Taylor expansion of complex scalar fields of
the usual infinitesimal chiral diffeomorphisms induces an algebraic
framework, which, embedded into a B.R.S. setting, leads to another
presentation of $\cW$-algebras (Eq.\rf{brsK}) written in terms of jet-ghosts.

(iv) due to physical considerations require to transform these ghosts
from jets into tensors. Obviously, this change of generators is not unique.

In doing so, we have given a general solution (for any order $s$ of
the algebra), which put into the game all the truncation mechanisms of
the Forsyth frames up to order $s$ via a differential order reduction
(DOR).  The price to pay in keeping the entire generality of the
solution is to carry  the weight of the whole hierarchy  of
differential equations (with orders lower than $s$) which rule all the
linear truncations.  However, if one considers, for a given level,
the general solution \rf{Walgebra} as (physically) uncompleted,  the
removing of the role of the  intermediate levels to the benefit of the
standard $\cW$-algebra presentation comes as a satisfying surprise.
It has been shown that the cancellation holds  in a rather tractable
way for the lowest orders and can be related with some known
computation \cite{Gov96}. However,  the existence of non trivial
redefinitions of the $\cC \lra \tld{\cC}$ ghosts leading  to the
re-absorption  of the intermediate DOR decompositions,  could be an
very interesting problem. This gives a sharper indication on the
nature of the tensorial ghosts $\tld{\cC}$ associated to infinitesimal
$\cW$-diffeomorphisms. It is a close issue to the one concerning the
relationship  between the parameters of the KdV flows and those of
infinitesimal $\cW$-diffeomorphisms \cite{Gov96}. In particular, how
the nested variations pertaining to the various sub-levels are finally
disentangled to the benefit of the top level only, deserve to be
better studied. All comes from both the conformal covariance (geometry
and global meaning) and the nilpotency of the BRS operation
(associative algebra of  symmetry). This gives an algorithm similar to
one obtained in \cite{Gov96}, in which, conformal covariance governs
the calculation as well.

(v) $\cW$-currents are differential polynomials in the structure
functions $\cR\zbzs$ only.

Further, a window on the so-called $\cW$-gravity is open, once the BRS
algebra is given, with the use of the algebraic trick given in
\cite{Ba88,Bandelloni:1999et,BaLa02} in order to incorporate the
sources of the $\cW$-currents. The relationship between the
$\cW$-diffeomorphism symmetry and the Beltrami deformation parameters
for the complex geometry is given by
\begin{eqnarray*}
  \bprt = \big\lbrace \delta_{\cW_s},\frac{\prt}{\prt\bc}
  \big\rbrace,\qquad
 \qquad \rho^{(\ell)} =
\frac{\prt \tld{\cC}^{(\ell)}}{\prt\bc}\ ,
\end{eqnarray*}
where $\bc$ is the true diffeomorphism ghost along the direction
$\bprt$ and the $\rho^{(\ell)}$ are expected to be the sources for the
$\cW$-currents. This justifies \rf{dictionary} and allows to get the
whole BRS algebra for $\cW$-gravity directly from the BRS algebra for
$\cW$-algebra (e.g. \rf{W3alg} or \rf{W4alg}). In particular, this will
be useful for a
systematic study of $\cW$-anomalies possibly arising at the quantum level.

As a final conclusion, we emphasize once more  that,  due to the non
linearity of this type of symmetry algebra of large (chiral)
diffeomorphisms, the technical intricacy is 
just a consequence of the reduction from jets to tensors for which non
trivial explicit solutions have been obtained. The latter can be
considered as a starting point for   a more pleasant treatment, and a
more suitable physical formulation for general $\cW$-algebras and
their relationship not only with linear algebraic differential equations
\cite{Forsyth59,DiFrancesco:1991qr}, but also with some kind of
differential systems. For instance, one ought to expect
that the Bershadsky $\cW_3^{(2)}$-algebra be rather related to a
conformally covariant system of coupled differential equations (with as
unknowns $f$ and $g$) of the form \cite{Garajeu:1995jn}
\begin{eqnarray*}
  \big(\prt_{(2)} +  a_{(1)}\zbz \prt + a_{(2)}\zbz \big) f\zbz +
  b\zbz g\zbz &=& 0 \\[2mm] 
\big(\prt - \sm{1}{2} a_{(1)}\zbz\big) g\zbz + B\zbz f\zbz &=& 0\ ,
\end{eqnarray*}
over a generic Riemann surface.

\paragraph{Acknowledgements.} We thank INFN-TS11 (Italy) and R\'egion
PACA (France) for their financial support to our collaboration.
Special thanks are due to the referee for several suggestions in improving the
version of the present work, in particular, in getting a more general
algorithm, this brought the subsection \ref{compar}. 





\end{document}